\documentclass[]{aa}  

\usepackage{graphicx}
\usepackage{txfonts}
\usepackage{soul}
\usepackage[colorlinks=true,allcolors=magenta]{hyperref}%
\begin{document}

   \title{XXL-HSC: The link between AGN activity and star formation in the Early Universe ($z\geqslant3.5$)}

    \titlerunning{The link between AGN activity and star formation in the Early Universe}
    \authorrunning{E. Pouliasis et al.}

    \author{E.~Pouliasis\inst{1}
    \and G.~Mountrichas\inst{2}
    \and I.~Georgantopoulos\inst{1}
    \and A.~Ruiz\inst{1}
    \and R.~Gilli\inst{3}
    \and E.~Koulouridis\inst{1}
    \and M.~Akiyama\inst{4}
    \and Y.~Ueda\inst{5}
    \and C.~Garrel\inst{6}
    \and T.~Nagao\inst{7}
    \and S.~Paltani\inst{8}     
    \and M.~Pierre\inst{6}
    \and Y.~Toba\inst{5,7,9,10}
    \and C.~Vignali\inst{11,3}
}

    \institute{IAASARS, National Observatory of Athens, Ioannou Metaxa and Vasileos Pavlou GR-15236, Athens, Greece\\
    \email{epouliasis@noa.gr}
    \and
    Instituto de Fisica de Cantabria (CSIC-Universidad de Cantabria), Avenida de los Castros, 39005 Santander, Spain
    \and
    INAF - Osservatorio di Astrofisica e Scienza dello Spazio di Bologna, Via Gobetti 93/3, I-40129 Bologna, Italy
    \and
    Astronomical Institute, Tohoku University, 6-3 Aramaki, Aoba-ku, Senda, 980-8578, Japan
    \and
    Department of Astronomy, Kyoto University, Kitashirakawa-Oiwake-cho, Sakyo-ku, Kyoto 606-8502, Japan
    \and 
    AIM, CEA, CNRS, Université Paris-Saclay, Université Paris Diderot, Sorbonne Paris Cité, F-91191 Gif-sur-Yvette, France
    \and
    Research Center for Space and Cosmic Evolution, Ehime University, 2-5 Bunkyo-cho, Matsuyama, Ehime 790-8577, Japan
    \and
    Department of Astronomy, University of Geneva, ch. d'Écogia 16, CH-1290 Versoix, Switzerland
    \and
    Academia Sinica Institute of Astronomy and Astrophysics, 11F of Astronomy-Mathematics Building, AS/NTU, No.1, Section 4, Roosevelt Road, Taipei 10617, Taiwan
    \and
    National Astronomical Observatory of Japan, 2-21-1 Osawa, Mitaka, Tokyo 181-8588, Japan
    \and
    Università di Bologna, Dip. di Fisica e Astronomia “A. Righi”, Via P. Gobetti 93/2, I-40129 Bologna, Italy}

   \date{Received  ; accepted }

 
  \abstract
  {
  
   {In this work, we aimed at investigating the star formation rate (SFR) of active galactic nuclei (AGNs) host galaxies in the early Universe.}
   {To this end, we constructed a sample of 149 luminous ($\rm L_{2-10keV} > 10^{44}\,erg\,s^{-1}$) X-ray AGNs at $\rm z \geq3.5$ selected in three fields with different depths and observed areas (Chandra COSMOS Legacy survey, XMM-{\it{XXL}} North and eROSITA Final Equatorial-Depth Survey). We built their spectral energy distributions (SED) using available multi-wavelength photometry from X-rays up to far-IR. Then, we estimated the stellar mass, M$_{*}$, and the SFR of the AGNs using the X-CIGALE SED fitting algorithm.}
  {After applying several quality criteria, we ended up with 89 high-z sources. More than half (55\%) of the X-ray sample have spectroscopic redshifts. Based on our analysis, our high-z X-ray AGNs live in galaxies with median $\rm M_{*}=5.6 \times10^{10}~M_\odot$ and $\rm SFR_{*}\approx240\,M_\odot yr^{-1}$. The majority of the high-z sources ($\sim89$\%) were found inside or above the main sequence (MS) of star-forming galaxies. Estimation of the normalised SFR, $\rm SFR_{NORM}$, defined as the ratio of the SFR of AGNs to the SFR of MS galaxies, showed that the SFR of AGNs is enhanched by a factor of $\sim 1.8$ compared to non-AGN star-forming systems. Combining our results with previous studies at lower redshifts, we confirmed that $\rm SFR_{NORM}$ does not evolve with redshift. Using the specific black hole accretion rate (i.e., $\rm L_X$ divided by $\rm M_{*}$), $\rm \lambda _{BHAR}$, that can be used as a tracer of the Eddington ratio, we found that the bulk of AGNs that lie inside or above the MS have higher specific accretion rates compared to sources below the MS. Finally, we found indications that the SFR of the most massive AGN host galaxies ($\rm log\,(M_{*}/ M_\odot) >10^{11.5-12}$) remains roughly constant as a function of M$_*$, in agreement with the SFR of MS star-forming galaxies.}

  }


   \maketitle
%
\section{Introduction}

The bulk of massive galaxies and also a fraction of lower mass galaxies in the local Universe host in their centre a supermassive black hole \citep[SMBH,][]{magorrian1998,kormendy2004,filippenko2003,barth2004,greene2004,greene2007,dong2007,greene2008}. When extragalactic gas or gas originating in the host galaxy accretes into the SMBH, a huge amount of energy is released across the whole electromagnetic spectrum. This is characteristic of active galactic nuclei (AGNs). During the last decades, there is accumulating evidence that there is a connection between host galaxy evolution and black-hole growth. Indeed, using observational data, the star formation in galaxies presents a peak at the "cosmic noon" (at redshifts between $\rm z=1-3$) and decreases rapidly down to $\rm z=0$ in a similar way to the black hole accretion rate \citep{dickinson2003,hopkins2007,delvecchio2014}. Moreover, there is a strong correlation between the black hole mass and the host galaxy properties, such as the mass of the bulge \citep{magorrian1998,mclure2002} and the velocity dispersion \citep{ferrarese2000,gebhardt2000,kormendy2013}. However, the mechanisms lying behind these relations are not well understood and a scenario of AGN power and galaxy co-evolution is still in debate. 

SMBHs are triggered by accretion of cold gas. This cold gas also sets off the star formation of galaxies. Therefore, many studies support the idea that the presence of an AGN suppresses the star formation, either by depleting the available cold gas or through AGN feedback, i.e. via winds \citep{hopkins2016, bieri2017} or via relativistic jets \citep{heckman2014}. Nevertheless, AGN and star formation may co-exist during a galaxy evolution phase \citep[e.g.][]{Koulouridis2006a,Koulouridis2014}. This implies that either there is a common fuelling mechanism \citep[e.g., galaxy mergers;][]{hopkins2008, Bower2006, Koulouridis2006b,Koulouridis2013} or that AGNs provides positive feedback to their host, by, e.g., over-compressing cold gas through outflows \citep[e.g.,][]{zubovas2013}.

Based on observations in the nearby and early Universe, there is a conflict between different studies. In the low redshift regime, \citet{shimizu2017} used ultra-hard X-ray-selected AGNs from the {\it{Swift Burst Alert Telescope}} (BAT) at $\rm z<0.05$ and found that AGNs present enhanced star formation rate (SFR) compared to a control galaxy (non-AGN) sample. Based on their analysis, the SFR of AGNs shows a small dependence on AGN luminosity with no evidence for an upturn at high luminosities. \citet{leslie2016} used SDSS sources at $\rm z<0.1$ and classified them based on their  emission line ratios. They found that AGN feedback suppresses the star formation of the host galaxy. Caution has to be taken, though, when comparing results from different studies. AGNs constitute a diverse population, and different selection criteria choose AGN with different properties \citep[e.g., $\rm L_X$,][]{pouliasis2019,pouliasis2020}. The analysis followed to estimate galaxy properties (e.g., SFR) may also lead to systematic biases \citep{mountrichasBOOTES}. Furthermore, it should also be taken into account that galaxy control samples that have different properties (e.g., control sample with only star-forming galaxies versus a simple mass-matching control sample) may lead to apparent different conclusions when compare their SFRs with that of AGN systems \citep{shimizu2017}.

At higher redshifts, \citet{santini2012} used X-ray AGNs from three fields (GOODS-North, GOODS-South and COSMOS) and compared their SFRs with that of a mass-matched control sample of non-AGN galaxies, at $\rm 0.5<z<2.5$. They found that AGNs have enhanced far infrared (FIR) emission compared to non-AGN systems with similar mass. \citet{mahoro2017} studying a sample of FIR detected AGNs and non-AGN systems at z$\sim$0.8 in the COSMOS field found a positive AGN feedback. \citet{florez2020} used X-ray-selected AGNs in Stripe 82 with $\rm L_{2-10keV} > 10^{44}\,erg\,s^{-1}$ and compared their SFRs with non-X-ray galaxies at $\rm 0.5<z<3$. Based on their analysis, AGNs present 3-10 times enhanced SFR compared to non-X-ray systems. \citet{mountrichasBOOTES, mountrichas2022, mountrichasEFEDS} used X-ray AGNs in the  Bo$\rm \ddot{o}$tes, COSMOS and eFEDS fields, respectively, and compared their SFRs with that from reference galaxy control samples. They found that at $\rm L_{2-10keV} < 10^{44}\,erg\,s^{-1}$ X-ray AGNs tend to have lower SFRs than that of star forming main sequence \citep[MS,][]{brinchmann2004,noeske2007, whitaker2012, speagle2014,schreiber2015} galaxies, while at higher luminosities AGNs present enhanced SFRs, at least  for systems within a specific stellar mass, M$_*$, range ($\rm 10.5 < log\,(M_*/M_\odot) < 11.5$).

Since at high redshifts ($\rm z>1$), galaxy (non-AGN) samples are limited in size, a number of studies that compared the SFR of X-ray AGNs with that of non-AGN systems, used for the latter analytical expressions from the literature. Towards this end, they estimated the normalised SFR, $\rm SFR_{NORM}$. $\rm SFR_{NORM}$ is defined as the ratio of the SFR of an AGN over the SFR of a star-forming MS galaxy with similar stellar mass and redshift as the AGN. For the latter, the analytical equation 9 of \citet{schreiber2015} is often used. These studies found that $\rm SFR_{NORM}$ is independent of redshift \citep{mullaney2015}, that the $\rm SFR_{NORM}$ distribution of higher $\rm L_X$ AGNs is narrower and shifted to higher values compared to their lower L$_X$ counterparts \citep{bernhard2019}, that the effect of AGN on the SFR of the host galaxy depends on the location of the galaxy relative to the MS \citep{masoura2018} and that there is a strong correlation between the $\rm SFR_{NORM}$ and the X-ray luminosity, with lower $\rm L_X$ AGNs to lie below the MS and higher $\rm L_X$ AGNs residing in host galaxies above the MS \citep{masoura2021}.

The aforementioned studies were focused on X-ray selected AGN samples up to $\rm z\leq 3$ . In this work, we aim at examining the position of X-ray AGNs relative to the MS in the early Universe ($\rm z \geq3.5$). To this end, we construct a sample of high-z sources spanning a wide range of X-ray luminosities in the bright end ($\rm L_{2-10keV} > 10^{44}\,erg\,s^{-1}$). In particular, we use high-z sources selected in three fields with different areas and depths (CCLS, XMM-XXL and eFEDS). We construct the spectral energy distributions (SEDs) of the sources to derive their host galaxies properties, such as the M$_*$ and the SFR, using the X-CIGALE SED fitting algorithm. Our goal is to study the SFR of the most luminous X-ray AGNs that can be detected by these surveys, at high redshifts, and compare them to the MS star-forming galaxies. Since, at such high redshifts, non-AGN galaxies are scarce, we follow the approach of previous studies at lower redshifts \citep[e.g.][]{bernhard2019, masoura2021} and estimate the $\rm SFR_{NORM}$ parameter using the analytical expression from \cite{schreiber2015}.

The construction of the high-z sample is presented in Sect.~\ref{data}. In Sect.~\ref{analysis}, we derive the host galaxy properties using SED fitting and we consider the reliability of our sample. In Sect.~\ref{results}, we discuss the derived properties of the host galaxies and we compare them with the MS of star-forming galaxies. Then, we examine the position of X-ray AGNs with respect to the MS. In Sect.~\ref{summary}, we summarise the results. Throughout the paper, we assume a $\Lambda$CDM cosmology with $\rm H\textsubscript{0}=70$ km s\textsuperscript{-1} Mpc\textsuperscript{-1}, $\rm \Omega\textsubscript{M}=0.3$ and $\rm \Omega \textsubscript{$\Lambda$}=0.7$.

\section{Sample}\label{data}

\begin{table*}
\caption{Summary of the photometry available in each field and their corresponding average limiting magnitudes. The depths are referred to the 5$\sigma$ limiting magnitude in AB system. }              
\label{numbercounts}      
\centering                                      
\begin{tabular}{l c c c c }          
\hline\hline                        
Survey   &  \multicolumn{3}{c}{Depth (5$\sigma$ AB)} \\
 Filters       & CCLS  &  XMM-XXL  &  eFEDS & References \\

\hline                                  
GALEX:FUV,NUV &  $\sim$26,$\sim$25.5  & $\sim$22.7 & 19.9, 20.8 & 1,19,20\\
\\
HSC:g,r,i,z,y &  27.3,26.9,26.7,26.3,25.3  & 26.6,26.2,26.2,25.3,24.5 & 26.6,26.2,26.2,25.3,24.5 & 2\\
KiDS: u,g,r,i &  --  & -- & 24.2, 25.1, 25.0, 23.7 & 3\\
LS8: g,r,z &  --  & -- & 24.0, 23.4, 22.5 &4 \\
CFHTLS: u,g,r,i,z &  26.3,26.0,25.6,25.4,25.0  & 26.3,26.0,25.6,25.4,25.0 & -- & 5  \\
\\
CFHT/WIRDS: J,H,K &  23.4,24,24  &  -- & -- &  21 \\

VISTA/UltraVISTA: Y,J,H,Ks&  25,24,24,24  &  -- & -- &  22 \\

VISTA/VHS: J,Ks &  --  &  21.1,19.9 & 21.1,19.9 & 9\\
VISTA/VIKING: z,J,H,Ks   & --   &23.1,22.3,22.1,21.5,21.2 &  23.1,22.3,22.1,21.5,21.2 & 10,11\\
VISTA/VIDEO: Y,J,H,Ks   & --   &24.5,24.4,24.1,23.8 &  -- & 12\\
UKIDSS/LAS: Y,J,H,K &  20.8,20.5,20.2,20.1  &  -- & -- &  6 \\

UKIDSS/UDS: J,H,K &  --  &  24.9,24.7,24.9 & -- &  6,7 \\
UKIDSS/DXS: J,K &  --  &  $\sim$22.6,$\sim$22.1 & -- &  6,8 \\

VIPERS/MLS: Ks   & --   &$\sim$22 &  -- & 13\\
\\
Spitzer/IRAC: 1,2,3,4 &  24.0,23.3,21.3,21.0  & 22.1,21.5,19.65,19.5&-- &14,15\\
WISE: 1,2,3,4 &  --  & 19.2,18.8,16.4,14.5 & 21.0,20.1,16.7,14.5 & 16,17\\

Spitzer/MIPS: 1    &  19.3  & 21.4&-- & 15,18 \\
\\
Herschel/PACS: 1,2   & 14.7,13.9   &13.65,13.3 & 13.4,12.7& 15,18 \\

Herschel/SPIRE: 1,2,3   &  14.1,13.8,13.4  & 14.9,14.9,14.9& 14.9,14.9,14.5& 15,18 \\

\hline  
\end{tabular}\label{depth}
\tablebib{
(1)~\citet{bianchi2014}; (2)~\citet{aihara2019}; (3)~\citet{kuijken2019}; (4)~\citet{dey2019}; (5)~\citet{hudelot2012}; (6)~\citet{lawrence2007UKIDDS}; (7)~\citet{almaini2007}; (8)~\citet{swinbank2013}; (9)~\citet{mcmahon2013}; (10)~\citet{edge2013}; (11)~\citet{kuijken2019}; (12)~\citet{jarvis2013}; (13)~\citet{moutard2016}; (14)~\citet{vaccari2015}; (15)~\citet{laigle2016}; (16)~\citet{meisner2019}; (17)~\citet{cutri2012}; (18)~\citet{shirley2021}; (19)~\citet{zamojski2007}; (20)~\citet{capak2007}; (21)~\citet{bielby2012}; (22)~\citet{mccracken2012}.
}
\end{table*}

\begin{table*}
\caption{Number of the high-z sources used in our analysis before and after applying the quality criteria (QS, Sect.~\ref{criteria}) and percentages of available photometry in each wavelength window. MIRS and MIRL stand for the shortest (3.4-4.6 $\rm \mu m$) and longest (22-24 $\rm \mu m$) MIR bands, respectively.}              
\label{numbercounts}      
\centering                                      
\begin{tabular}{c | c | c c c c c  | c | c c c c c }          
\hline\hline                        
  &    \multicolumn{6}{c |}{Before QC} & \multicolumn{6}{c}{After QC}\\
  \hline 
 Field  & Total & Spec-z & NIR & MIRS & MIRL & FIR & Total & Spec-z & NIR & MIRS & MIRL & FIR \\
   & number & \multicolumn{5}{c |}{Percentage (\%)} & number & \multicolumn{5}{c}{Percentage (\%)} \\

\hline                                  
   CCLS   & 53 & 47 & 64 & 77 & 77 &  77   &  28& 50     & 79 & 100 & 100 & 100  \\
    XMM-XXL   & 54 & 52  & 91 & 78 & 50 &   37  &  39 & 44   & 95 & 82 & 49 & 39  \\
    eFEDS     & 42 & 57  & 45 & 100 & 79 & 26 &  22 & 82     & 73 & 100 & 73 & 41  \\ 
\hline 
    Total    &149 &  52  & 69 & 84 & 68 &  48 & 89 & 55      & 85 & 92 & 71 & 60  \\ 
\hline  
\end{tabular}\label{tableNumber}
\end{table*}

In this section, we give a brief description of the high-redshift samples used in this work and their available photometry. In this work, we used X-ray catalogues with sources in the early Universe ($z\geqslant3.5$). We constructed the sample using data from different fields of various areas and depths to compile a dataset with the highest possible completeness with respect to the high end of the luminosity and the redshift ranges. We used the X-ray high-z sources selected in the CCLS, XMM-XXL and eFEDS fields:

\begin{itemize}

\item The Chandra COSMOS Legacy Survey \citep[CCLS,][]{civano2016} covers an area of 2.2 $\rm deg^2$ and includes Chandra observations of about 4.6\,Ms, reaching a depth of $\rm 2.2\times 10^{-16} erg~cm^{-2}~s^{-1}$ in the soft X-ray band (0.5-2 keV). \citet{marchesi2016} provided the optical and infrared identifications for the whole sample of 4016 X-ray sources in the CCLS field and obtained the photometric redshifts using the LePhare code \citep{arnouts1999, ilbert2006}. There are 53 sources with $z\geqslant3.5$. 25 out of 53 (47\%) have available spectroscopic redshifts.

\item In the XMM-XXL North field \citep[][XXL Paper I]{pierreXXL} that covers an area of about 25 $\rm deg^2$ at a depth of $\rm \sim 6 \times 10^{-15}\,erg~cm^{-2}~s^{-1}$ (at 3$\rm \sigma$) in the soft band (0.5-2 keV), we used the high-z sample presented in \citet[][]{pouliasis2021}. The X-ray data used in the latter study rely on an internal release obtained with the V4.2 XXL pipeline. This sample was selected using the Hyper Suprime-Cam \citep[HSC,][]{miyazaki2018} colour-colour diagrams and verified using the X-CIGALE fitting algorithm for the redshift estimation. The initial catalogue contains 91 high-z sources. Out of these, 28 have spectroscopic redshifts, while 63 have photometric redshifts. In our analysis, we only selected sources with secure redshift higher than 3.5. Specifically, we used all the 28 sources with spectroscopic redshifts $\rm z \geq3.5$ and 26 out of 63 sources whose photometric redshift probability density functions, PDF(z), peaks at a value higher than 3.5. Therefore, our XXL sample consists of 54 high-z sources. 52\% of these have spectroscopic redshifts.

\item We used data obtained with the Extended ROentgen Survey with an Imaging Telescope Array \citep[eRosita,][]{predehl2021}, onboard the Spektrum-Roentgen-Gamma mission. Specifically, we used the X-ray catalogue in the eROSITA Final Equatorial-Depth survey \citep[eFEDS,][]{brunner2021} field. eFEDS covers an area of 140 $\rm deg^2$ with an average exposure time of $\rm \sim 2.2 ~ks$ ($\rm \sim1.2 ~ks$ after correcting for telescope vignetting) that corresponds to a limiting flux of $\rm F_{0.5-2 ~keV} \sim 7 \times 10^{-15}~ erg ~s^{-1} ~cm^{-2}$. \citet{salvato2021} built a catalogue with the multi-wavelength information of the X-ray sources and their redshift estimations. We excluded sources with low reliability with respect to their photometric redshift measurement (CTP\_REDSHIFT\_GRADE $\leq 3$) and/or sources whose the assignment of the counterpart is unreliable (CTP\_quality<2). This resulted in 42 sources with $\rm z \geq3.5$. 24 out of 42 ($\sim$57\%) sources have spectroscopic redshifts. 
\end{itemize}

By combining the three fields, we obtained a total of 149 X-ray-selected AGNs with $z\geqslant3.5$ and $\rm L_{2-10keV} > 10^{44}\,erg\,s^{-1}$. 77 out of 149 ($\sim$52\%) have spectroscopic redshifts (Table~\ref{tableNumber}). The identification of the multi-wavelength counterparts of our high-z X-ray sources is described in
\citet{marchesi2016}, \citet{pouliasis2021} and \citet{salvato2021}, for the COSMOS, XXL and eFEDS fields, respectively. The photometry is complemented using the publicly available catalogue of the Herschel Extragalactic Legacy Project \citep[HELP,][]{shirley2019,shirley2021}. HELP catalogue combines observations across the entire electromagnetic spectrum over the Herschel Multitiered Extragalactic Survey \citep[HerMES,][]{oliver2012} and the H-ATLAS survey \citep{eales2010} covering the ultraviolet (UV) to FIR part of the spectrum. Using HELP enriches our sample with multi-wavelength data  in addition to FIR information obtained with the Photodetector Array Camera and Spectrograph \citep[PACS,][]{poglitsch2010} and the Spectral and Photometric Imaging Receiver \citep[SPIRE,][]{griffin2010}. In Table~\ref{depth}, we list the deepest surveys used in our analysis in the three fields. We also provide the average limiting magnitudes (5$\sigma$ in AB system) in each filter.

We cross-matched our high-z samples with the HELP catalogue using the optical coordinates of our X-ray sources and a search radius of 1". To assess the reliability of the cross-matching method, we calculated the probability that a match is associated to the true object instead of being a random projection in the sky. Thus, we generated catalogues with random positions from the three X-ray samples by shifting the original positions between $\pm 1^{\circ}$ along the right ascension. Then, we calculated the false positive matches and, following the formula of reliability defined as $\rm R=1-N_{shifted}^{matches}/N_{original}^{matches}$, we ended up with an average reliability larger than 97\% for the three samples. Out of the 149 high-z sources all are detected in the optical filters. 68.5\% of them have also available NIR photometry. 83.9\% and 67.8\% of the sources are detected in the shortest and longest MIR bands, respectively, while 48.3\% have been detected in either PACS or SPIRE (or both) observations. Table~\ref{tableNumber} summarises these numbers also for the individual fields.


\begin{figure}
    \begin{tabular}{c}
     \includegraphics[width=0.47\textwidth]{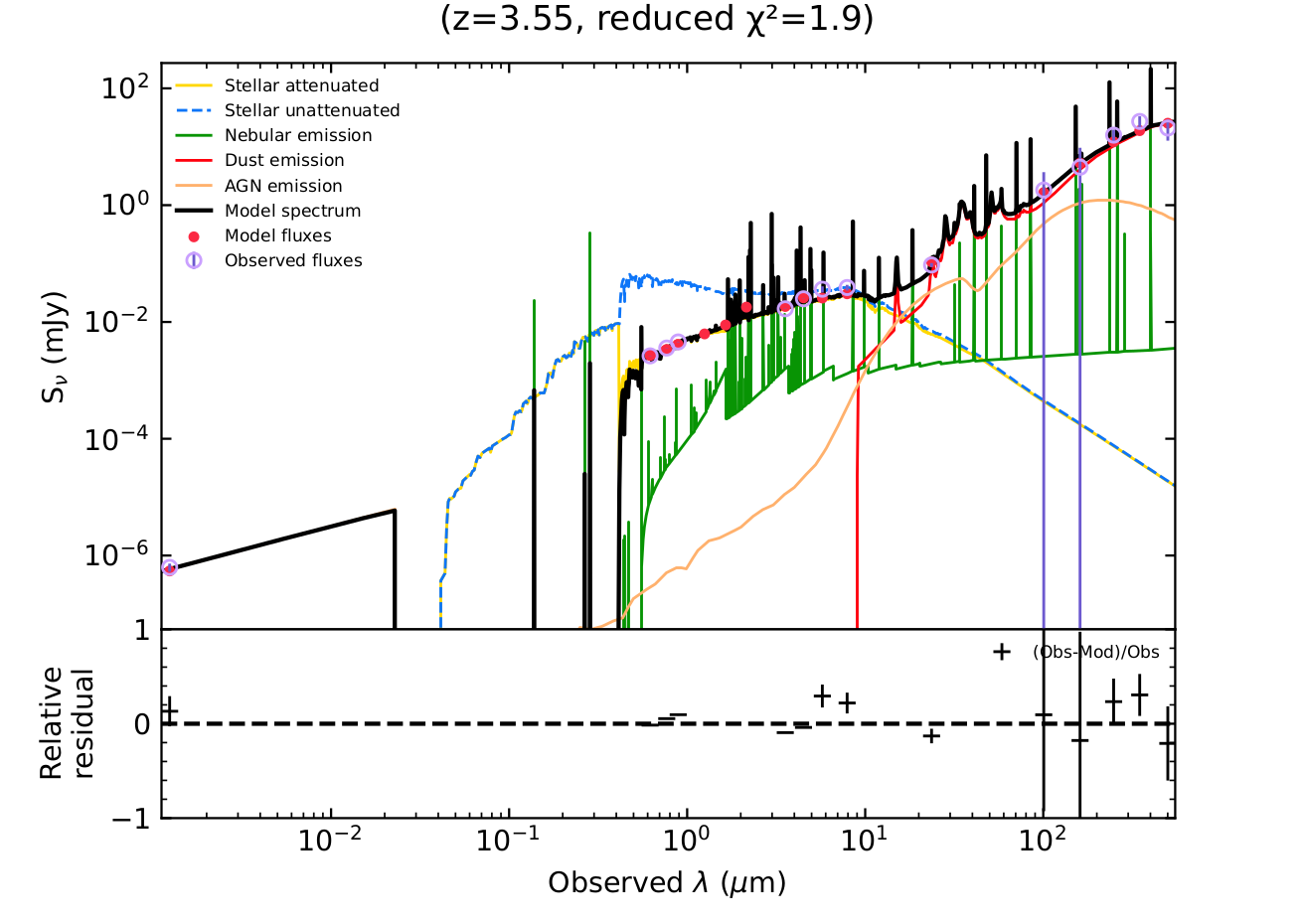} \\
     \includegraphics[width=0.47\textwidth]{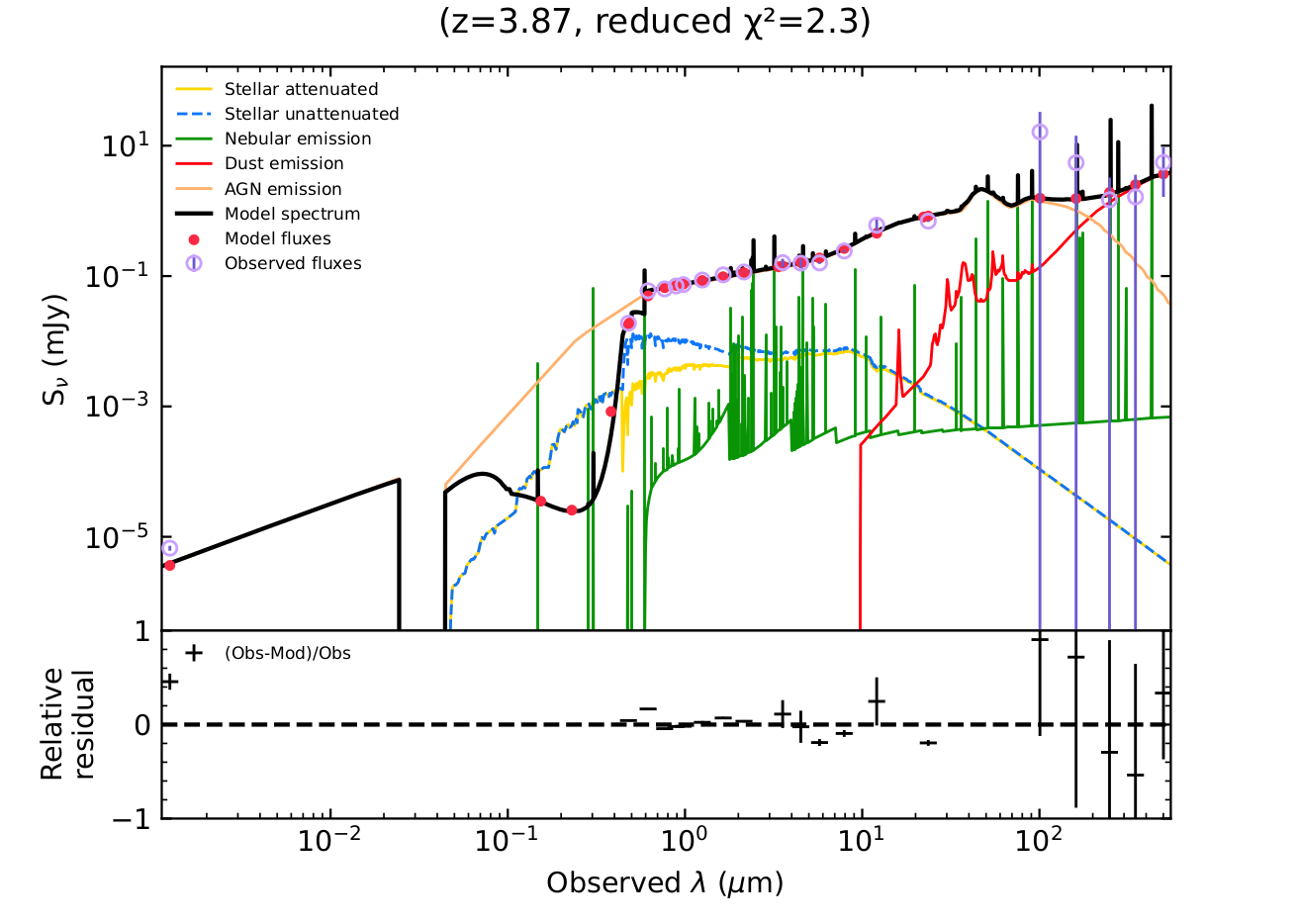}\\
     \includegraphics[width=0.47\textwidth]{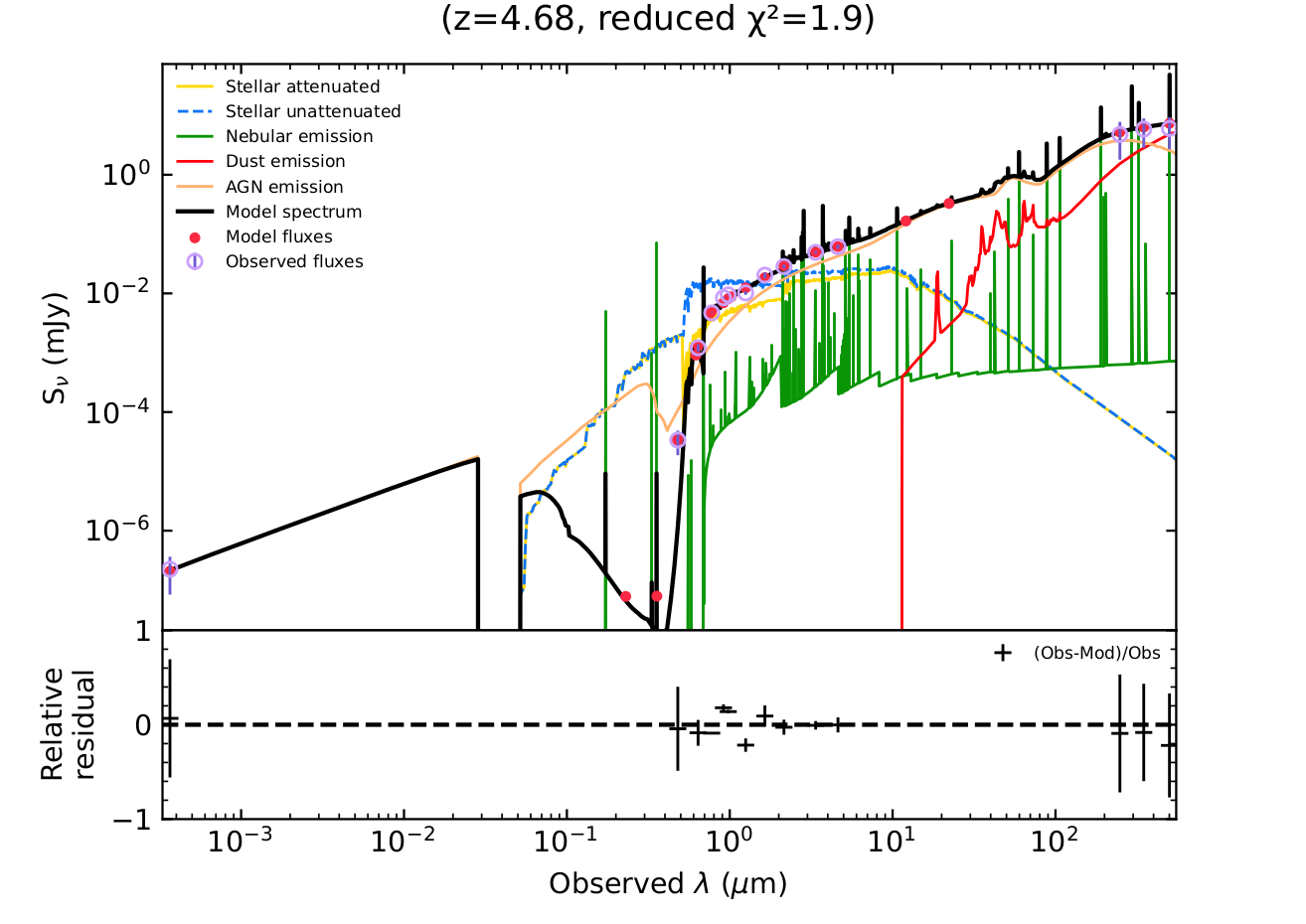}
    \end{tabular}
\caption{Examples of SEDs in the three fields CCLS, XMM-XXL and eFEDS (from top to bottom). The dust emission is plotted in red, the AGN component in green, the attenuated (unattenuated) stellar component is shown with the yellow (blue) solid (dashed) line, while the orange lines shows the nebular emission. The total flux is represented with black colour. Below each SED, we plot the relative residual fluxes versus the wavelength.}\label{exampleSED}
\end{figure}


\begin{table*}
\caption{Models and their parameter space used by \texttt{X-CIGALE} for the SED fitting of the high-z sources.}
\begin{tabular}{ l c r }
\hline
\multicolumn{1}{l}{Parameter} &  & Value \\ \hline \hline
\multicolumn{3}{c}{Star formation history: delayed SFH with optional exponential burst)}\\
Age of the main stellar population in Myr && 500, 1000, 1500, 2000, 3000, 4000, 5000 \\
e-folding time of the main stellar population model in Myr, $\tau_{\rm main}$ && 200, 500, 700, 1000, 2000, 3000, 4000, 5000 \\
Age of the late burst in Myr, $age_{\rm burst}$ && 20  \\
Mass fraction of the late burst population, $f_{\rm burst}$ && 0.0, 0.005, 0.01, 0.015, \\
&& 0.02, 0.05, 0.10, 0.15, 0.18, 0.20\\
e-folding time of the late starburst population model in Myr, $\tau_{\rm burst}$ && 50.0\\
\hline
\multicolumn{3}{c}{Stellar population synthesis model}\\
Single Stellar Population Library&&\citet{bruzual2003}\\
Initial Mass Function&& \citet{chabrier2003} \\
Metallicity && 0.02 (Solar) \\
\hline
\multicolumn{3}{c}{Nebular emission}\\
Ionization parameter ($\log U$)&& -2.0 \\
Fraction of Lyman continuum escaping the galaxy ($f_{\rm esc}$)&& 0.0 \\
Fraction of Lyman continuum absorbed by dust ($f_{\rm dust}$)&& 0.0 \\
Line width (FWHM) in km/s&& 300.0 \\
\hline
\multicolumn{3}{c}{Dust attenuation: modified attenuation law \citet{charlot2000}} \\
V-band attenuation in the interstellar medium, $Av_{ISM}$ && 0.2, 0.3, 0.4, 0.5, 0.6, 0.7, \\
&& 0.8, 0.9, 1, 1.5, 2, 2.5, 3, 3.5, 4\\
\hline
\multicolumn{3}{c}{Dust template: \citet{dale2014}}\\
AGN fraction && 0.0\\
Alpha slope, $\alpha$ && 2.0\\
\hline
\multicolumn{3}{c}{AGN models from \citet{stalevski2016} (SKIRTOR)}\\
 Average edge-on optical depth at 9.7 micron (t) &&   3.0, 7.0\\
 Power-law exponent that sets radial gradient of dust density (pl) && 1.0\\
 Index that sets dust density gradient with polar angle (q) && 1.0 \\
 Angle measured between the equatorial plane and edge of the torus (oa) && 40\\
 Ratio of outer to inner radius, $\rm R_{\rm out}/R_{\rm in}$ && 20\\
 Fraction of total dust mass inside clumps ($\rm M_{\rm cl}$) && 97\%\\
 Inclination angle ($i$) && 30, 70\\
 AGN fraction && 0.0, 0.1, 0.2, 0.3, 0.4, 0.5, 0.6, 0.7, 0.8, 0.9, 0.99 \\
 Extinction in polar direction, E(B-V) && 0.0, 0.2, 0.4\\
 Emissivity of the polar dust && 1.6 \\
 Temperature of the polar dust (K) && 100.0\\
 The extinction law of polar dust && SMC\\
\hline
\multicolumn{3}{c}{X-ray module}\\
Photon index ($\Gamma$) of the AGN intrinsic X-ray spectrum && 1.8 (CCLS), 2.0 (XXL \& eFEDS) \\
Maximum deviation of $\alpha_{ox}$, max\_dev\_$\alpha_{ox}$ && 0.2\\
LMXB photon index && 1.56 \\
HMXB photon index && 2.0 \\
\hline
\end{tabular}
\tablefoot{Edge-on, type-2 AGNs have inclination $i=70$~degrees and face-on, type-1 AGNs have $i=30^\circ$. The extinction in polar direction, E(B-V), included in the AGN module, accounts for the possible extinction in type-1 AGNs, due to polar dust. The AGN fraction is measured as the AGN emission relative to IR luminosity (1--1000 $\mu$m).}
\label{proposal}  
\end{table*}

\section{Data Analysis}\label{analysis}

In this section, we describe the SED fitting analysis we followed to derive the host galaxy properties of the X-ray AGNs in our sample. Moreover, we describe the criteria we applied to select only sources with robust measurements.

\subsection{SED fitting and parameter estimation}\label{sed}

We used the \texttt{X-CIGALE} code \citep{yang2020,yang2022} to derive the physical properties of the high-z sources and specifically the M$_*$ and the SFR. \texttt{X-CIGALE}, is a new branch of the Code Investigating GALaxy Emission \citep[\texttt{CIGALE},][]{boquien2019}, a multi-component SED fitting algorithm that fits the observational data of the sources to the theoretical models, and has been widely used in the literature \citep[e.g][]{pouliasis2020, padilla2021,toba2021a, toba2021b,toba2021c,yang2021}. X-CIGALE offers new capabilities, such as the inclusion of the X-ray absorption-corrected flux, $\rm f_X$, in the fitting process and accounts for extinction of the UV and optical emission in the poles of AGNs \citep{yang2018, mountrichasXXL, buat2021,toba2021a}. In Appendix~\ref{xrays}, we present how we obtained the intrinsic $\rm f_X$ for each target field. We used all the available photometry to construct the SEDs, from X-rays up to FIR, and we created a grid that models both the galaxy and the AGN emission. In particular, we used the stellar population synthesis model of \citet{bruzual2003} assuming the initial mass function (IMF) by \citet{chabrier2003} and constant solar metallicity (Z\,=\,0.02). We used the dust emission templates by \citet{dale2014} without AGN emission and the dust extinction law by \citet{charlot2000}. Furthermore, we adopted the AGN templates presented in \citet[SKIRTOR]{stalevski2012,stalevski2016} that are based on a realistic two-phase clumpy torus model, while for the star formation history (SFH) we used a delayed SFH with the functional form $\rm SFR\propto t\times exp(-t/\tau)$ that includes a star formation burst no longer than $\rm \tau = 20\,Myr$. The full list of the models and their parameters used in our analysis are given in Table~\ref{proposal}. With this configuration, we were able to fit the observational data with more than 300 million models. In Fig.~\ref{exampleSED} we show three SED examples representative of the objects of our samples used in this work. 

\begin{figure}
\center
   \begin{tabular}{c}
    \includegraphics[width=0.45\textwidth]{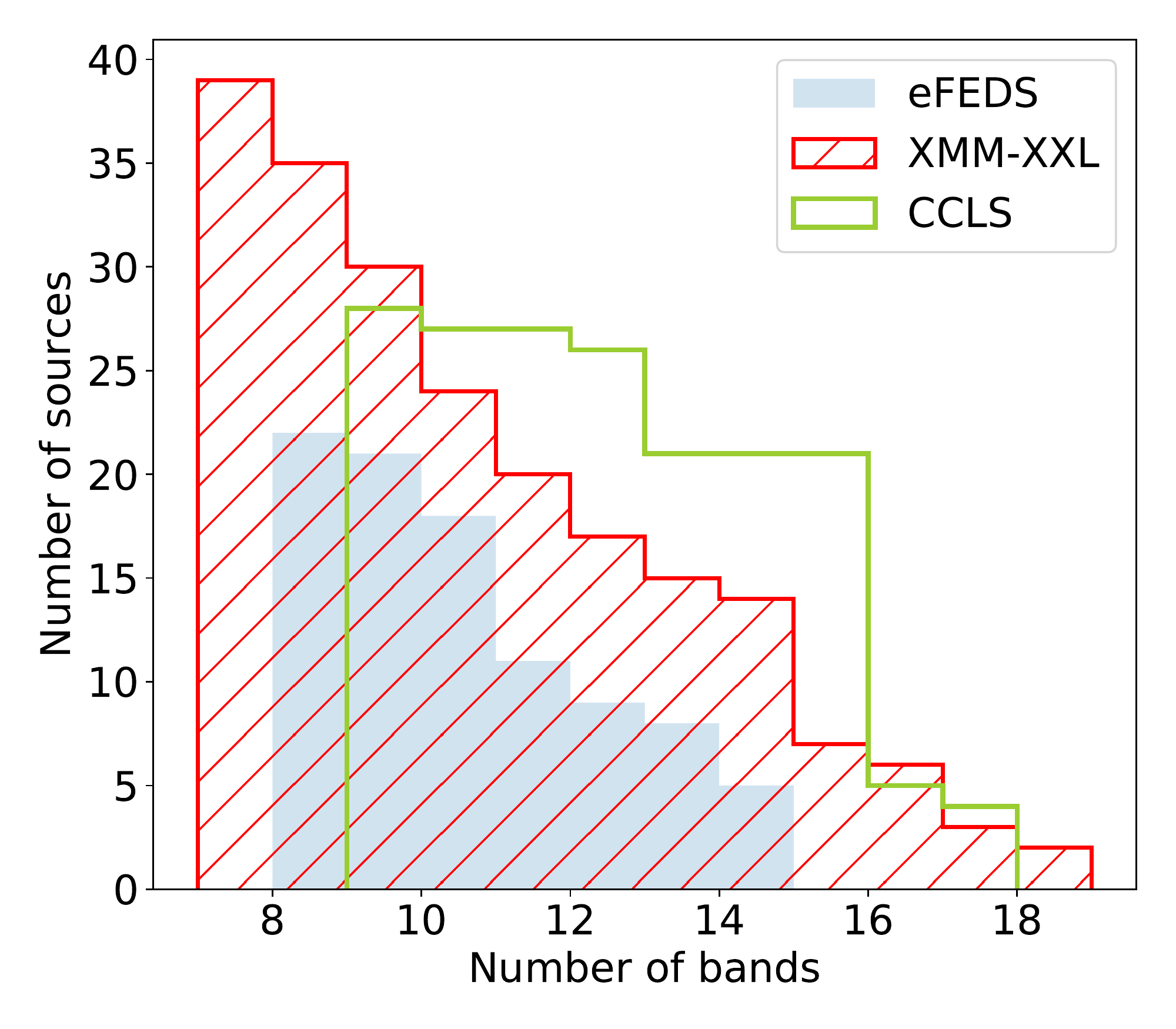} \\
    \end{tabular}
\caption{Cumulative number of sources versus the number of bands that a source has been detected. Different histograms represent the three target fields as indicated in the legend.}\label{nbands}
\end{figure}

\begin{figure}
\center
   \begin{tabular}{c}
    \includegraphics[width=0.45\textwidth]{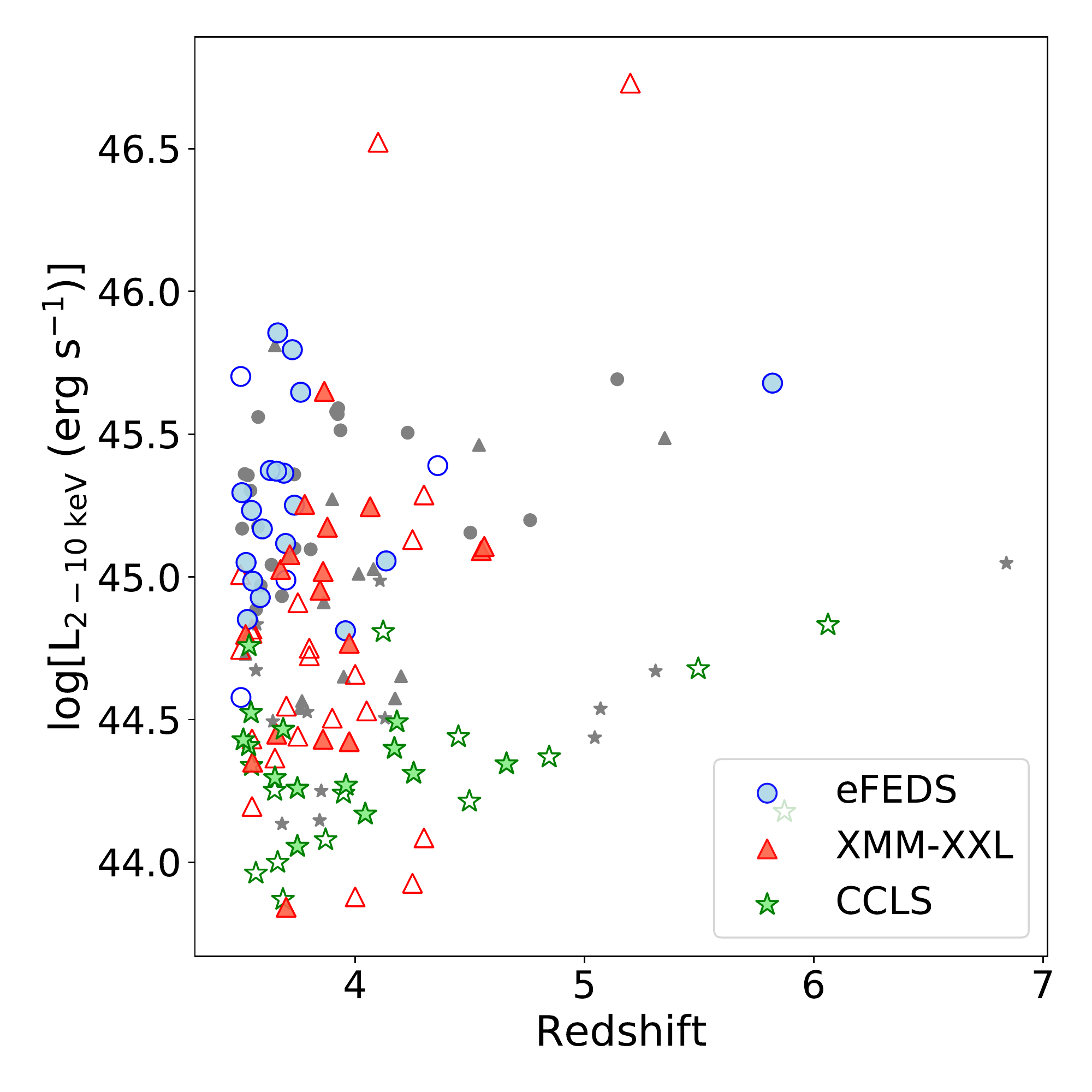} 
    \end{tabular}
\caption{X-ray absorption-corrected, rest-frame luminosity versus redshift for the X-ray sources detected in our three fields as indicated in the legend. The filled (empty) points correspond to sources with spectroscopic (photometric) redshift estimations. The smaller gray points represent those sources that were rejected after applying the quality selection criteria.}\label{fig_lx_distrib_3fields}
\end{figure}


\subsection{Quality selection criteria}\label{criteria}

It is important in our analysis, to have reliable measurements of both the global M$_*$ and the SFR of the AGN host galaxies. For that purpose, first we required our sources to have low reduced $\rm \chi^2$ ($\rm \chi^2_r$) that is indicative of the goodness of the SED fitting process. Therefore, we excluded sources that have $\rm \chi^2_r\geqslant5$ \citep[e.g.,][]{masoura2018,mountrichas2019,buat2021}. Furthermore, we applied quality criteria to the $M_*$ and SFR estimated parameters. X-CIGALE calculates for each of the parameters the best values and the Bayesian estimations with the corresponding uncertainties. The best values refer to the best-fit model, while the Bayesian takes into account the weights of all models. These weights are based on the likelihood with $\rm exp(-\chi^2/2)$ for each model \citep{boquien2019}. Large discrepancies may indicate unreliable parameter estimations. Following the recent studies of \citet{mountrichas2021,buat2021,koutoulidis2021}, we have included in our analysis only sources that satisfy the following two criteria: $\rm 1/5\leq M_{*,best}/M_{*,bayes}\leq 5$ and $\rm 1/5\leq SFR_{best}/SFR_{bayes}\leq 5$, where $\rm M_{*,best}$ and $\rm SFR_{best}$ refer to the best-fit values and the $\rm M_{*,bayes}$ and $\rm SFR_{bayes}$ to the Bayesian estimations. In the following of the paper, the Bayesian values of the properties are used unless it is specified otherwise.

After applying the quality selection criteria, we ended up with 89 high-z sources with secure host galaxy properties (28, 39 and 22 in CCLS, XMM-XXL and eFEDS, respectively). Regarding the sources that did not fulfill the above criteria, $\sim$33\% was not included in our analysis because of the $M_*$ and SFR criteria, while the remaining sources were excluded due to their high $\rm \chi^2_r$ values. The median number of bands per source for our final sample is 11, 11 and 15 for the eFEDS, XMM-XXL and CCLS samples, respectively. In Fig.~\ref{nbands}, we show the cumulative distribution of the number of sources as a function of the number of bands. All X-ray AGNs used in our analysis have optical photometry. 71 out of 89 sources have detections in the optical, near-infrared (NIR) and mid-infrared (MIR) bands. 75 out of 89 have NIR and 82 out of 89 have MIR photometric bands. There are no sources without both NIR and MIR photometry. 53 AGNs were detected by {\it{Herschel}}. Table~\ref{tableNumber} lists the number of high-z sources after applying the quality criteria. Comparing the photometric coverage to the initial sample, X-CIGALE and the quality selection criteria are able to reject sources that lack specific photometric bands.

\begin{figure*}
\center
   \begin{tabular}{c c}
    \includegraphics[width=0.45\textwidth]{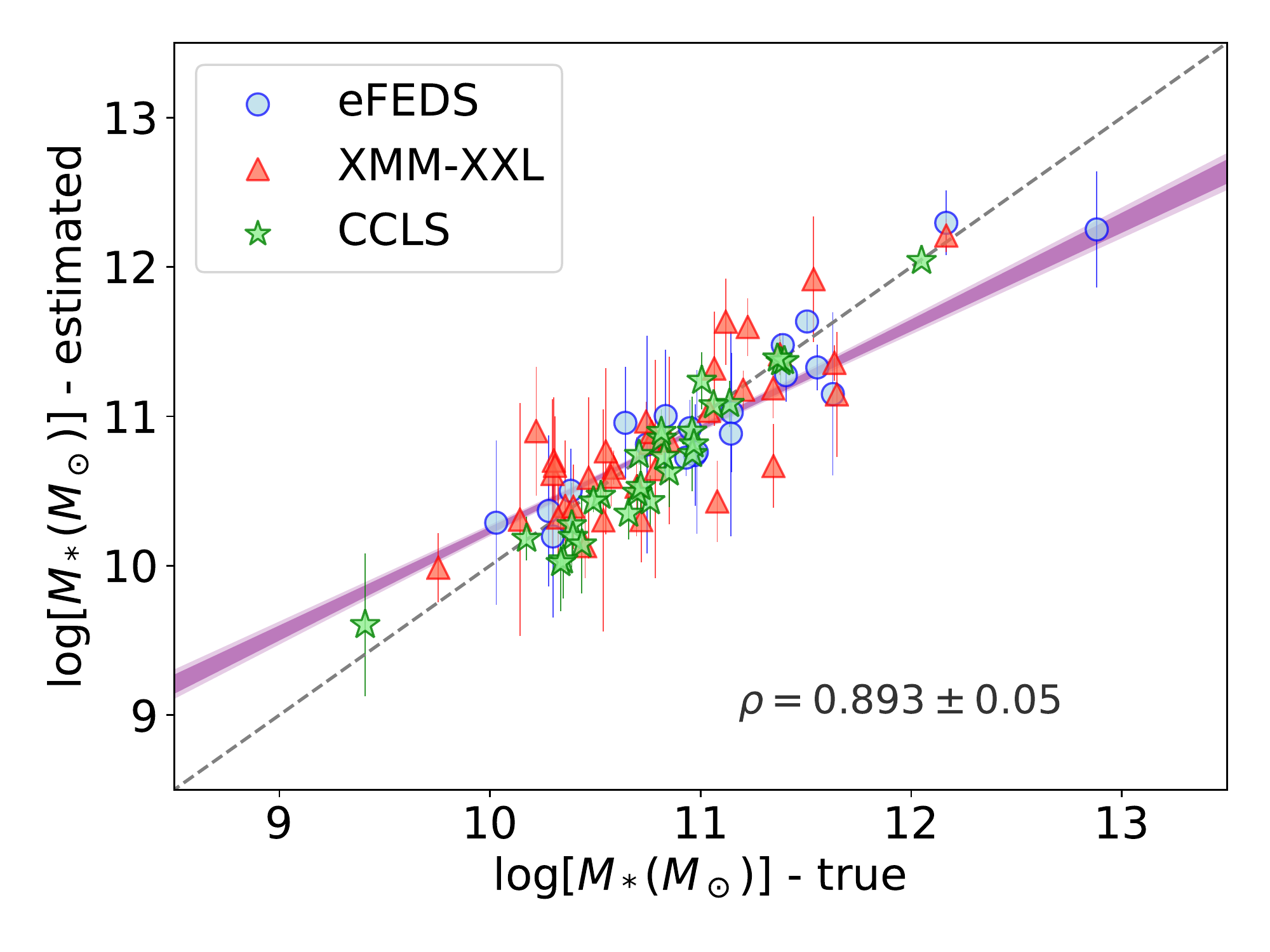}&
    \includegraphics[width=0.45\textwidth]{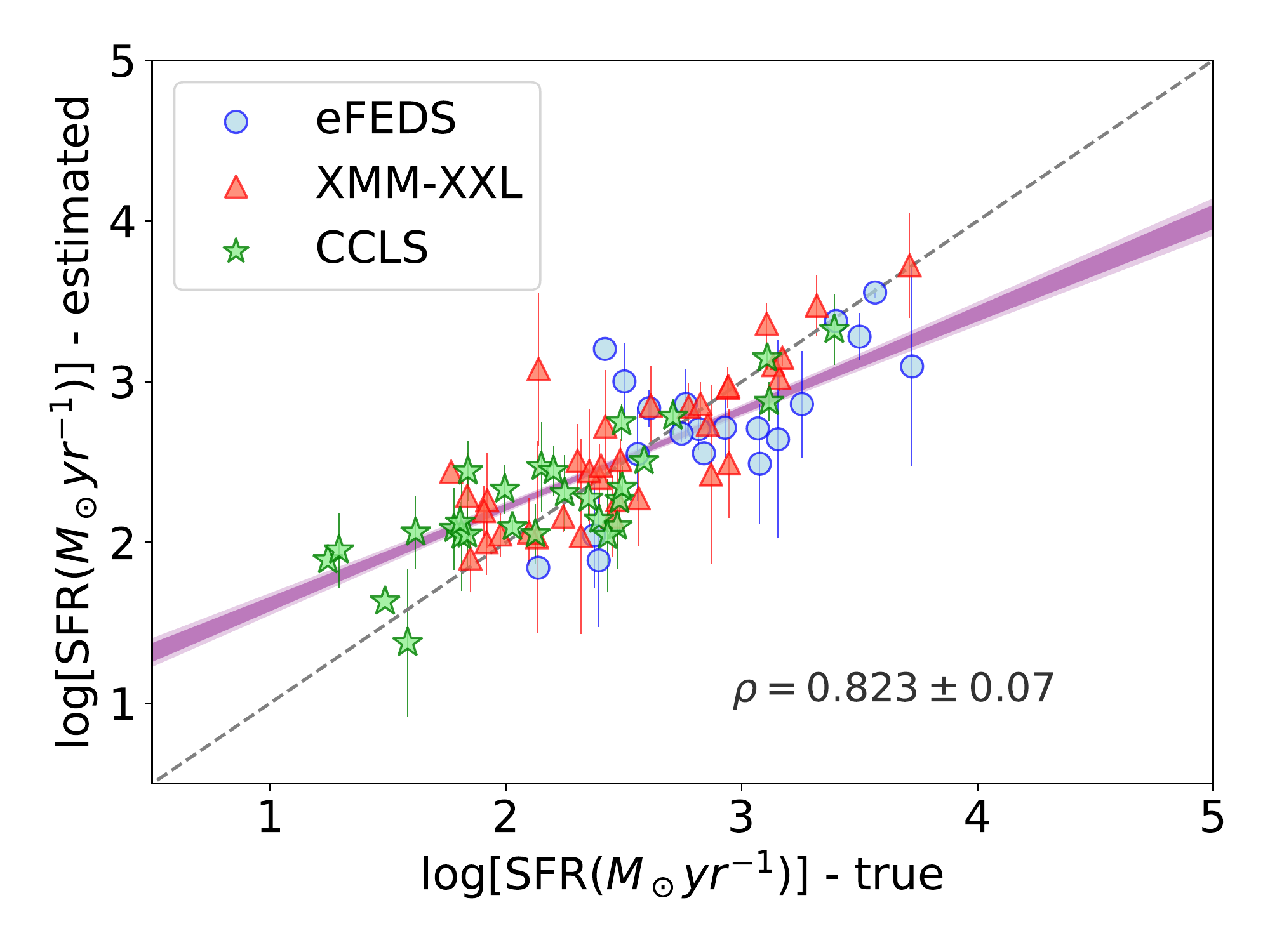} \\
    \includegraphics[width=0.45\textwidth]{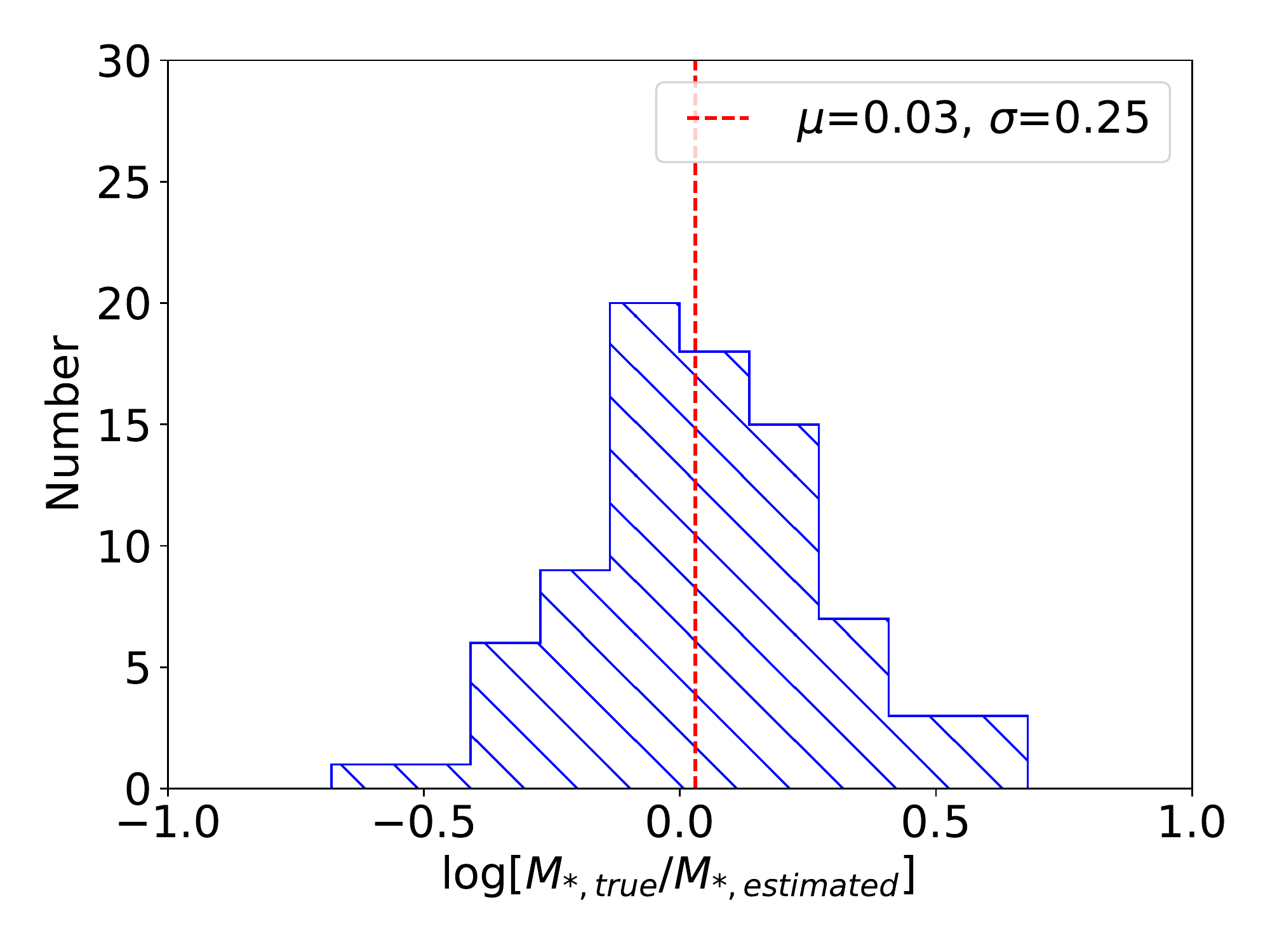} &
    \includegraphics[width=0.45\textwidth]{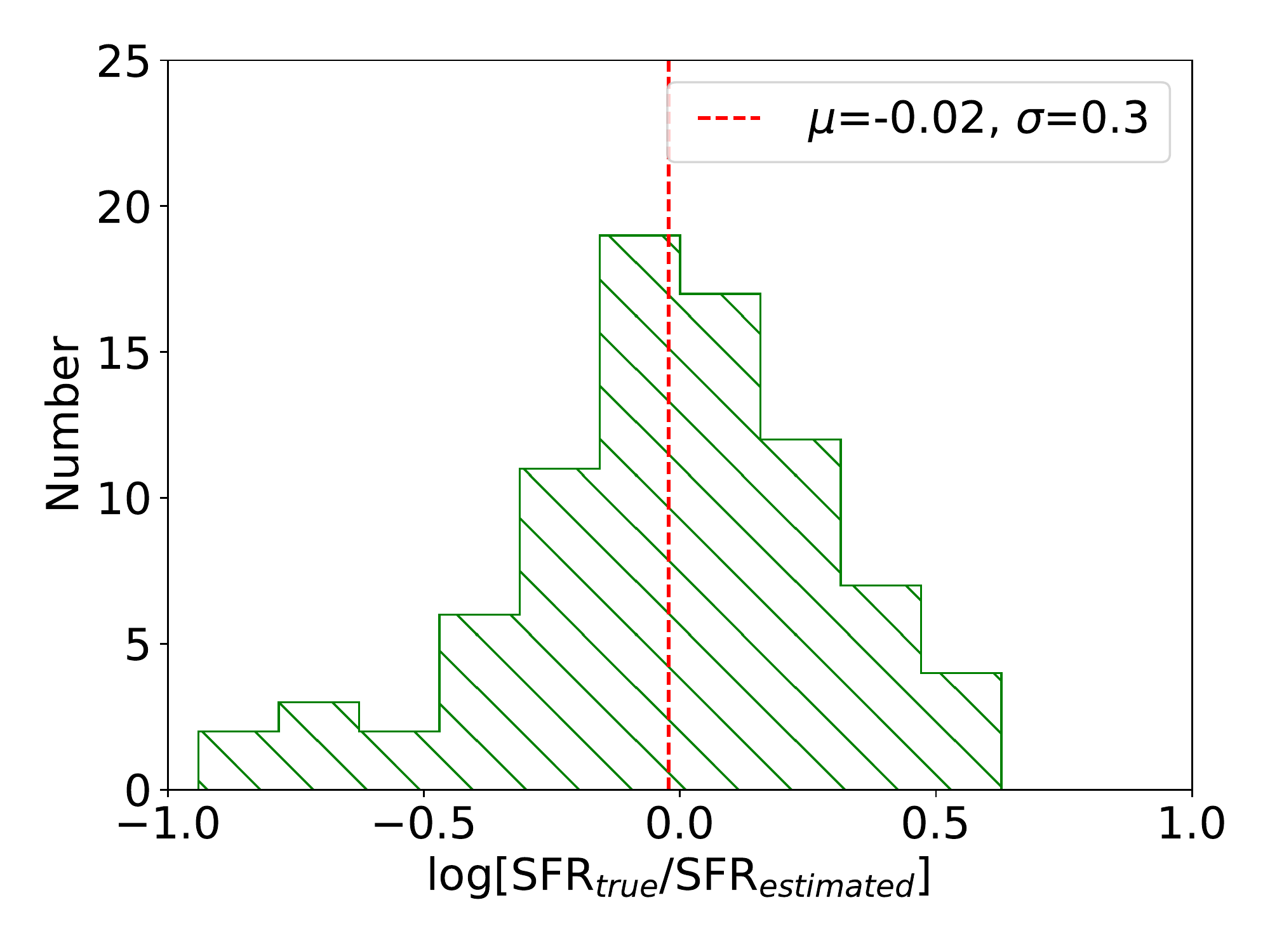}
    \end{tabular}
\caption{Upper panels: Estimated parameters, M$_*$ (left) and SFR (right) based on the mock catalogues compared to the true values provided by the best-fit model for the three fields used in our analysis as indicated. The dashed line represent the 1-to-1 relation to assist the plot interpretation, while in each plot we give the linear Pearson correlation coefficient ($\rm \rho$) of the combined sample. The shaded regions correspond to 1$\sigma$ and 2$\sigma$ confidence bands of the linear regression obtained with BCES. Lower panels: Corresponding distributions of the difference between estimated and true values. We provide also the mean and standard deviation in each case.}\label{mock_results}
\end{figure*}

Finally, among the 89 sources 55\% (49 sources) have spectroscopic redshifts. In Fig.~\ref{fig_lx_distrib_3fields}, we plot the X-ray rest-frame, intrinsic luminosity (as defined through the X-ray analysis and discussed in Appendix~\ref{xrays}) versus redshift for our final high-z sample in the three fields. The median values of $\rm L_X$ in each sample are given in Table~\ref{propertiesTable}. X-ray sources in eFEDS present on average the highest $\rm L_X$ among the three fields. However, there are two XMM-XXL sources with very high luminosities. These two sources have photometric redshift estimations and are considered as highly obscured AGNs ($\rm N_H \geq$23) according to the X-ray spectral analysis. Furthermore, there are a few XMM-XXL data points having lower $\rm L_X$ than the CCLS sources. These sources lie in the XMM-Spitzer Extragalactic Representative Volume Survey (XMM-SERVS) that has an area of 5.3 $\rm deg^2$ and exposure time of $\sim$46 ks \citep{chen2018}. Thus, their faint luminosities may be explained by the wide range of sensitivity in the XMM-XXL field due to these new observations performed after 2012. This can also be shown in Fig.~8 of \citet{pouliasis2021} where the positions of the high-z sources in XMM-XXL field in the [X-ray flux, optical magnitude] plane cover the area of sources detected in CCLS.

\subsection{Reliability of the parameter estimations}\label{sed}

To examine the reliability of the physical parameters estimated with the SED fitting, we performed a mock analysis, which is implemented in X-CIGALE. This method was suggested by \citet{giovannoli2011} and has been used widely in the literature \citep{2015A&A...576A..10C,boquien2019,mountrichas2021,toba2019,toba2020}. To create the mock catalogue, X-CIGALE uses the best-fit model for each of the sources and its corresponding physical properties and integrates it over the bands available in the observed data set. Then, adding a Gaussian noise to the fluxes (with $\sigma$ coming from the initial uncertainty of the observed fluxes) it creates the mock catalogue. In Fig.~\ref{mock_results}, we show the one-to-one relation of the estimated and true values of the parameters used in this work (SFR, M$_*$). We used the linear Pearson correlation coefficient, $\rho$, to test how well X-CIGALE is able to provide unbiased parameters using the specific sample. The estimates of the M$_*$ and SFR are well constrained with $\rm \rho=0.893 \pm 0.05$ and $\rm \rho=0.823 \pm 0.07$, respectively. The uncertainties were calculated using a bootstrap method. Furthermore, we used the bivariate correlated errors and intrinsic scatter \citep[BCES,][]{akritas1996,nemmen2012} linear regression method to obtain the relation between the parameters. BCES takes into account the uncertainties of each variable and allows for intrinsic scatter. We obtained best-fit models $y=0.69 \pm 0.05 \times x +3.38 \pm 0.57$ and $y=0.60 \pm 0.05 \times x +1.01 \pm 0.13$ for M$_*$ and SFR, respectively. In the same figure, we show the difference between true and estimated values for each case (lower panel). The mean value of the difference in logarithmic scale is 0.03 dex for M$_*$ and -0.02 dex for SFR with standard deviations of 0.25 and 0.30 dex, respectively. The above statistical analysis shows that the estimated M$_*$ and SFR values are reliable without any significant systematic bias.

\begin{figure}
       \includegraphics[width=0.47\textwidth]{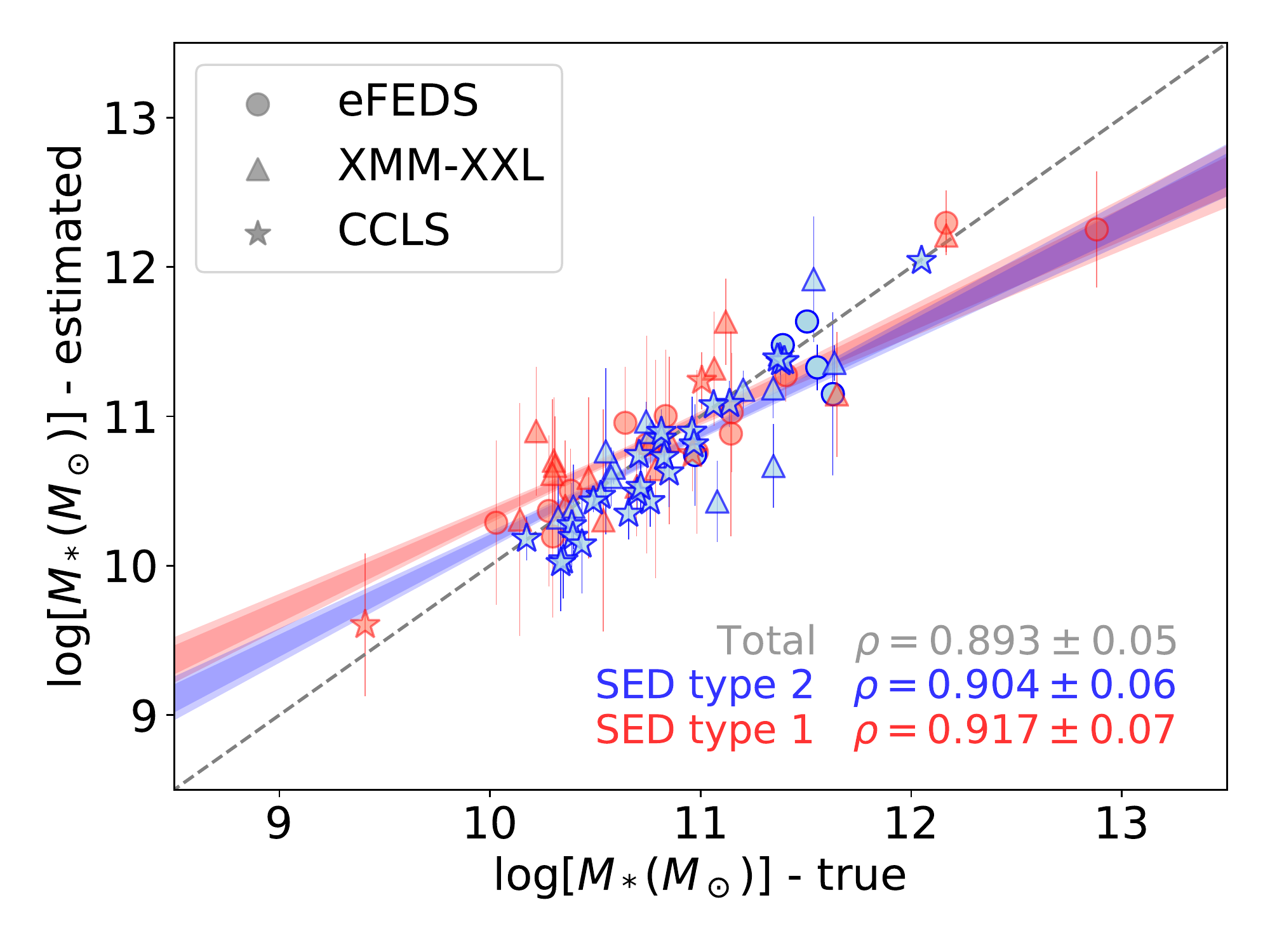} 
       \caption{Estimated M$_*$ based on the mock catalogue compared to the true values provided by the best-fit model for type 1 (red) and type 2 (blue) AGNs. The dashed line represent the 1-to-1 relation. The Pearson correlation factors for the two samples are $\rm \rho_{type 1}=0.917$ and $\rm \rho_{type2}=0.904$, respectively. The shaded regions correspond to 1$\sigma$ and 2$\sigma$ confidence bands of the linear regression obtained with BCES for the two samples.}\label{sedtype_mock}
\end{figure}

Another concern regarding the reliability of the derived properties is whether the SED fitting method is able to constrain M$_*$ for type 1 AGNs. This arises from the fact that in type 1 AGNs the optical to NIR part of the spectrum may be dominated by the AGN component and hence the host galaxy properties might be affected. To examine this, we explored whether there are any systematic differences in M$_*$ estimations between type 1 and type 2 AGNs. To this end, we used the results from the mock catalogues and compared the Pearson correlation factors derived separately for the type 1 and 2 AGN samples. In Fig.~\ref{sedtype_mock}, we plot the estimated M$_*$ versus the true values for the two samples. The best linear-fit functions for type 1 and type 2 AGNs are $y=0.68 \pm 0.08 \times x +3.56 \pm 0.85$ and $y=0.70 \pm 0.07 \times x +3.07 \pm 0.85$, respectively. In both cases, we found very strong correlations between the parameters ($\rm \rho_{type 1}=0.917$ and $\rm \rho_{type2}=0.904$). This means that X-CIGALE has the ability to derive secure M$_*$ estimations even when the AGN emission dominates that of the host galaxy, as in type 1 AGNs. However, we note that these results have been derived using a minimum of seven bands (Fig.~\ref{nbands}) in the SED fitting and may are not valid when lower number of bands are used. To classify the high-z sources into type 1 and type 2 AGNs according to the SED fitting, we used the Bayesian and the best-fit model values of the inclination angle as described in \citet{mountrichasXXL}. Secure type 1 sources are classified if $i_{Best} = 30^\circ$ and $i_{Bayes} < 40^\circ$ and secure type 2 if $i_{Best} = 70^\circ$ and $i_{Bayes} > 60^\circ$. 81 out of the 89 (91\%) high-z sources have secure type classification from X-CIGALE. Out of them, 36 and 45 sources were classified as type 1 and type 2, respectively.        

Moreover, in our analysis we used a constant metallicity (solar) for all of our sources. This prevents long time consuming calculations, but also it does not affect significantly the shape of the SEDs compared to the observed ones and the derived properties \citep{yuan2018,hunt2019}. However, since the metallicity evolves with redshift and maybe affect the AGN host galaxy properties estimations, we re-run X-CIGALE using a lower metallicity ($\rm Z=0.008$) value than the solar one. By comparing the results of the two runs, we found no significant differences. In particular, the $\rm \chi^2_r$ of the sources are similar in the two runs, while when comparing the M$_*$ and SFR estimations in the two runs we found a strong correlation between them ($\rm \rho=0.945$ and $\rm \rho=0.920$, respectively).

Finally, the reliability of the physical parameters estimates is sensitive to the photometric bands used to construct and fit the SEDs. M$_*$ is calculated based on the optical and NIR photometry, while FIR along with optical photometry is used by X-CIGALE to estimate the SFR parameter. Regarding the high-z sources, the shortest MIR bands cover the rest-frame NIR wavelengths assisting this way to constrain the M$_*$ parameter. Moreover, the longest MIR bands are required by X-CIGALE to constrain the AGN component at these redshifts. In Appendix~\ref{appendixB}, we examined whether the absence of some photometric bands affects the reliability of our SFR and M$_*$ calculations or the AGN physical properties. Based on our results, our measurements can be considered remarkably robust.

\section{Results}\label{results}

\subsection{Host galaxy properties}

\begin{figure}
\center
   \begin{tabular}{c}
    \includegraphics[width=0.47\textwidth]{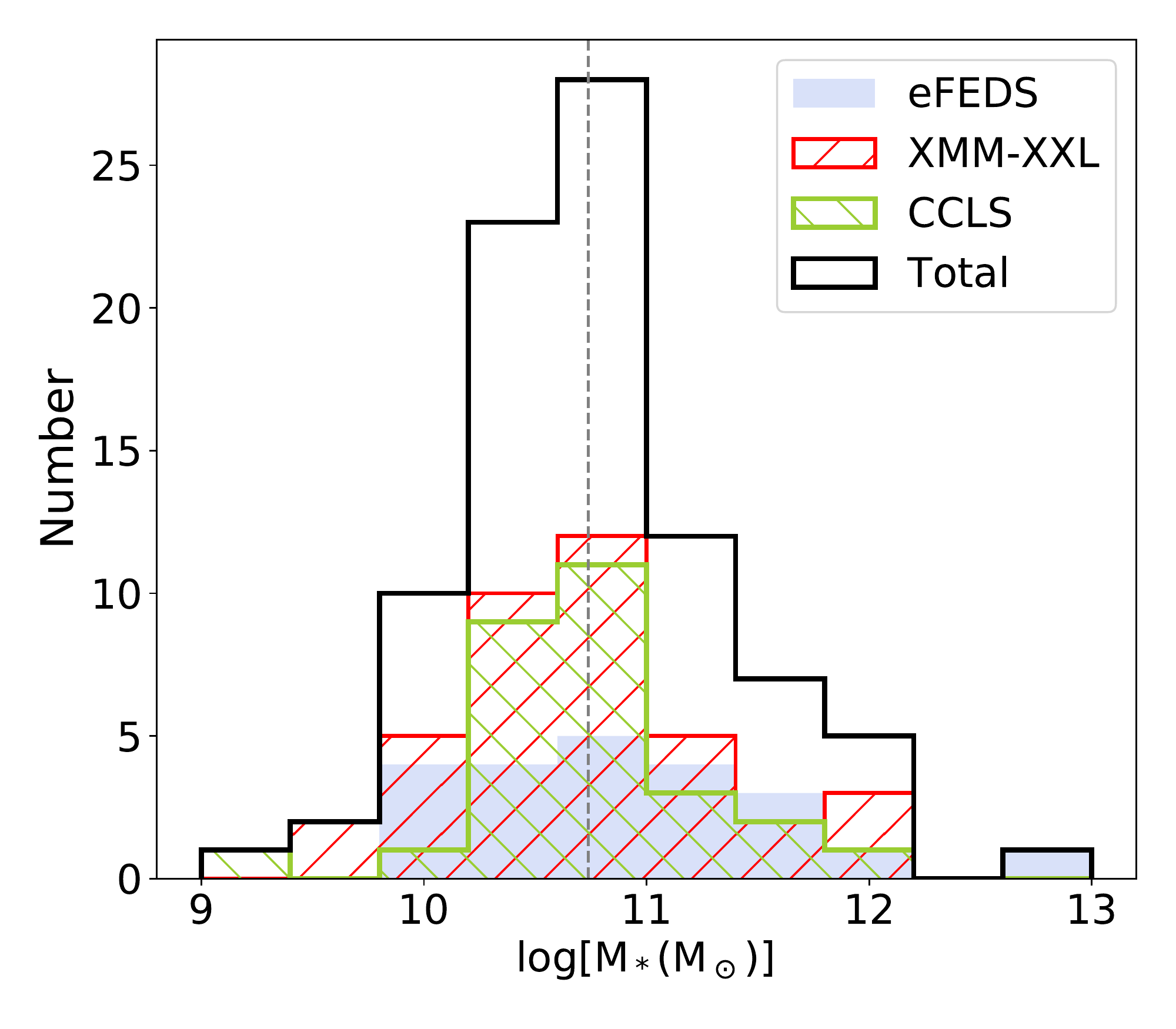} \\
    \includegraphics[width=0.47\textwidth]{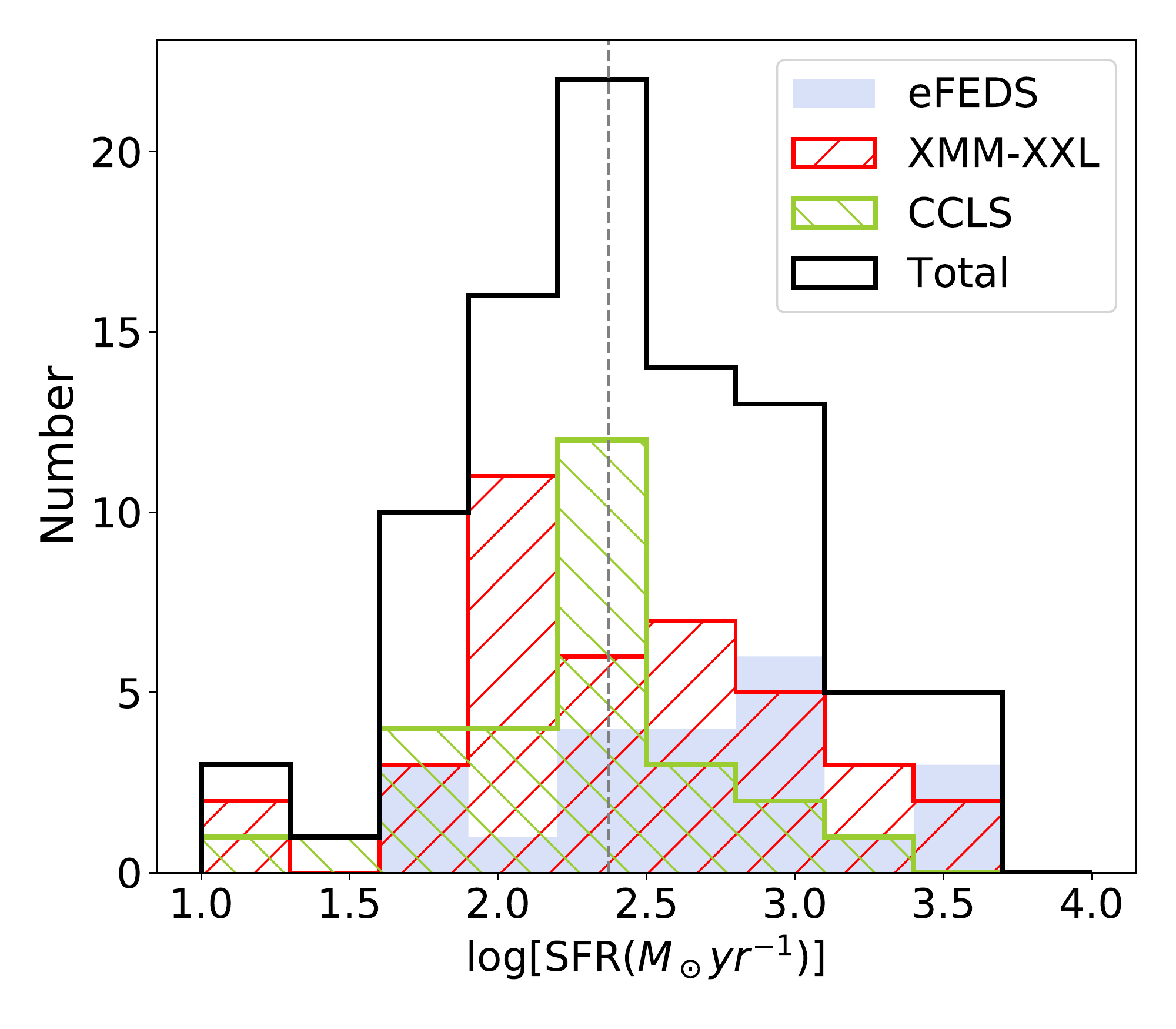}
    \end{tabular}
\caption{Distributions of M$_*$ (upper panel) and SFR (lower panel) for the high-z samples as indicated in the legend given in logarithmic scale. The dashed vertical lines correspond to the median values of the total sample.}
\label{properties}
\end{figure}

In this section, we present the derived properties of the high-z samples obtained using the SED fitting procedure and parameter space described in Sect.~\ref{sed}. In particular, for each field we calculate the median SFR and stellar mass, while we estimate the errors using a bootstrap resampling method \citep{loh2008}. The results are shown in Table~\ref{propertiesTable}. Figure~\ref{properties} shows the M$_*$ and SFR distributions of the high-z sources that met our criteria applied in Sect.~\ref{criteria} (well-fitted SED and secure estimated physical properties). In particular, we show the distributions of the total number of high-z sources (89) and the individual samples for each field (22, 39 and 28 objects in eFEDS, XMM-XXL and CCLS fields, respectively). The median values of the stellar mass for the three samples are $\rm med(M_{*,COSMOS})=5.7 \times10^{10} ~M_\odot$, $\rm med(M_{*,XXL})=4.2 \times10^{10} ~M_\odot$ and $\rm med(M_{*,eFEDS})=7.7 \times10^{10}~ M_\odot$  (Table~\ref{propertiesTable}). The AGNs from the eFEDS sample reside in slightly more massive galaxies. This might be expected, since eFEDS has an area of $\sim 140$ $\rm deg^2$ \citep{brunner2021} and thus it is possible to retrieve more luminous and massive galaxies. To evaluate if the three samples may come indeed from the same parent distribution, we performed a two-side Kolmogorov–Smirnov (K-S) test. The p-values are 0.78, 0.43 and 0.35 for the [XMM-XXL, CCLS], [XMM-XXL, eFEDS] and [eFEDS, CCLS], respectively, indicating that all distributions are similar. Combining the sources from the three fields, the estimated median is then $\rm M_{*,Total}=5.6 \times10^{10}~ M_\odot$. This value is consistent with that found in \cite{zou2019} at $\rm z>2$ using X-ray AGNs in the COSMOS field (see their Fig. 6).

The lower panel of Fig.~\ref{properties} presents the SFR distribution. This plot suggests moderate to high star forming activity. In XMM-XXL and CCLS fields, the median SFR is $\rm \sim 224\,M_\odot \, yr^{-1}$. AGNs in eFEDS appear to have slightly higher SFR, i.e., $\rm med(SFR_{*,eFEDS})=552\,M_\odot yr^{-1}$. Performing the K-S test between the three samples, only the difference between CCLS and eFEDS distributions shows statistical significance (p-value=0.0028). The median SFR in all three fields is $\rm SFR_{*,total}=236\,M_\odot yr^{-1}$. \cite{florez2020} used X-ray AGNs from Stripe 82X \citep{lamassa2013, lamassa2016}. For their highest redshift bin ($\rm z=2-3$), they found $\rm SFR=479\,M_\odot yr^{-1}$, which is in good agreement with our calculations. Our results are also in broad agreement with \cite{zou2019} that found $\rm SFR \sim 100\,M_\odot yr^{-1}$ for sources at $\rm z>2$ (see their Fig. 9). We note, however, that comparison of (host) galaxy's property measurements among different studies should be taken with caution since different systematic effects are introduced due to the different methodologies applied in various works for their estimation. Even in those cases where the same analysis has been applied (e.g. SED fitting), usage of different templates and parametric grid may introduce different systematic effects. This will be discussed further in Sect. \ref{sec_sfrnorm_lx}.


\subsection{Comparison of the SFR of AGNs with that of MS star-forming galaxies}

\begin{figure}
   \begin{tabular}{c}
       \includegraphics[width=0.48\textwidth]{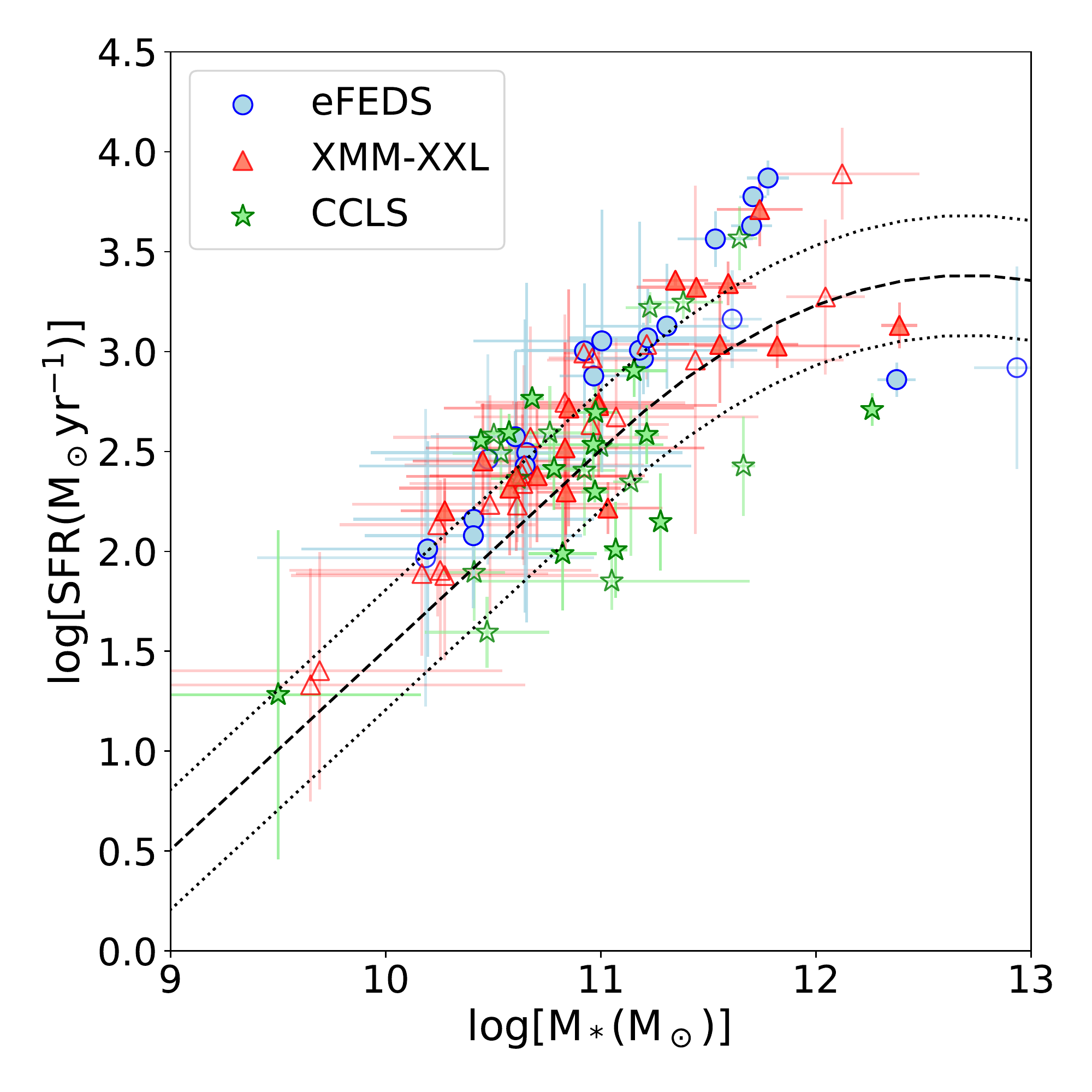} 
    \end{tabular}
\caption{SFR as a function of M$_*$. The different colours and shapes represent the samples used in our analysis as indicated in the legend. The dashed line represents the main sequence of star-forming galaxies obtained by \citet{schreiber2015} with median redshift value z=3.7. The dotted lines correspond to the uncertainties defined as $\pm0.3$ dex. The filled (empty) points correspond to sources with spectroscopic (photometric) redshift estimations.}\label{m_SFR}
\end{figure}

Star-forming galaxies show a tight correlation between their M$_*$ and SFRs, known as the MS of star-forming galaxies \citep[e.g., ][]{brinchmann2004,noeske2007, whitaker2012, speagle2014,schreiber2015}. This relation is valid through a wide redshift range and holds up to $\rm z \sim 4$ \citep{schreiber2016}. We used the derived properties of our high-z sample to examine the location of the X-ray AGNs with respect to the MS. In Fig.~\ref{m_SFR}, we plot our sources in the SFR-M$_*$ plane using the estimated SFR and M$_*$ values for our three samples. We compared their position relative to the MS, using for the latter the analytical expression of equation 9 of \citet{schreiber2015} (dashed line). For this calculation, we used the median redshift of our sample ($\rm z_{med}=3.7$). Furthermore, since \citet{schreiber2015} assumed a \citet{salpeter1955} IMF to derive this relation, while we adopted a \citet{chabrier2003} IMF, the $\rm M_*$ and SFR values have been re-scaled by applying a factor 1.63 and 1.59, respectively \citep{madau2014}.

The sources that have SFRs within $\rm 0.3$\,dex from the \citet{schreiber2015} SFR (dotted lines) were considered to lie within the MS, similarly to the criteria used in previous studies \citep[e.g.,][]{shimizu2015, koutoulidis2021}. In \citet{schreiber2015}, the SFR dispersion remains constant with redshift (z$=$0-4) at $\rm 0.3$\,dex, with the exception at the very low redshifts where there is an increase by $\rm \sim 0.4$\,dex. Moreover, \citet{matthee2019}, using the EAGLE cosmological hydrodynamic simulations, found the SFR scatter to decrease from z=0 up to z=5 by $\rm 0.05$\,dex. At similar redshifts and stellar masses to our study, they found a maximum dispersion of $\rm 0.25$\,dex. Finally, \citet{pearson2018} using deep Herschel data to derive the SFRs, found that the SFR dispersion decreases with redshift with a maximum of $\rm 0.15$\,dex at z$>$3.8. Hence, the SFR dispersion value adopted in our analysis is appropriate without overestimating the number of sources lying below or above the MS.

 Considering only the nominal values, our results revealed a large fraction of objects with enhanced star-forming activity compared to MS. Specifically, 41.6\% of the high-z AGNs lie above, 47.2\% inside and 11.2\% below the MS. Table \ref{propertiesTable} presents the percentages for the three fields individually. \cite{circosta2021}, as part of the SINFONI survey for Unveiling the Physics and Effect of Radiative feedback \citep[SUPER,][]{circosta2018}, used 27 X-ray AGNs with spectroscopic redshift $\rm 2.0<z<2.5$ from the COSMOS and XMM-XXL fields and estimated their host galaxy properties, by applying the SED fitting tool CIGALE. According to the results presented in their Fig. 1, $47.6\%$ of their AGNs lie above the MS \citep[defined by][]{schreiber2015}, $42.8\%$ are inside the MS and only $9.5\%$ are below the MS. Furthermore, \citet{vito2014b} using AGNs in the GOODS and COSMOS fields at z<1 also found that they are mostly hosted by star-forming galaxies. The results of the aforementioned studies are in very good agreement with our findings.

\begin{table*}
\caption{Properties of the high-z samples used in our analysis.}              
\centering                                      
\begin{tabular}{c c c c c c c c c}          
\hline\hline                       
Field & $\rm L_{2-10~keV}$ & M$_*$  & SFR & $\rm SFR_{NORM}$  &  $\rm \lambda_{sBHAR}$  & Above MS  & Inside MS & Below MS\\
      & $\rm log [erg~s^{-1}]$ & ($\rm 10^{10}~M_\odot$)  & ($\rm M_\odot~yr^{-1}$) &  log  &  $\rm \propto \lambda_{Edd}$  & (\%)  & (\%) & (\%)\\
 
\hline
   CCLS & 44.32 $\pm$ 0.05 &  5.7 $\pm$ 1.0  &  203 $\pm$ 31   & 0.07 $\pm$ 0.13  &   -0.87 $\pm$ 0.06  & 32.15 & 42.85 & 25.0 \\
    XMM-XXL & 44.77 $\pm$ 0.05 &  4.2 $\pm$ 1.2   & 235 $\pm$ 79   & 0.19 $\pm$ 0.01  &    -0.20 $\pm$ 0.10   & 35.89 & 61.54  & 2.56  \\
    eFEDS & 45.20 $\pm$ 0.09 &  7.7 $\pm$ 2.5 &  552 $\pm$ 91 & 0.33 $\pm$ 0.06  &   0.10 $\pm$ 0.20    & 63.63  & 27.27  & 9.09 \\ 
\hline 
    Total & 44.75 $\pm$ 0.13 & 5.6 $\pm$ 0.8   & 236 $\pm$ 32  & 0.20  $\pm$ 0.02 &     -0.47 $\pm$ 0.16    & 41.57 & 47.19 & 11.23    \\ 
\hline 
\end{tabular}
\tablefoot{The median values of the properties in each sample are presented. The last three columns present the percentages of the high-z X-ray AGNs that are located inside, above and below the main sequence of the star-forming galaxies.}\label{propertiesTable}  
\end{table*}

\begin{figure}
   \begin{tabular}{c}
       \includegraphics[width=0.47\textwidth]{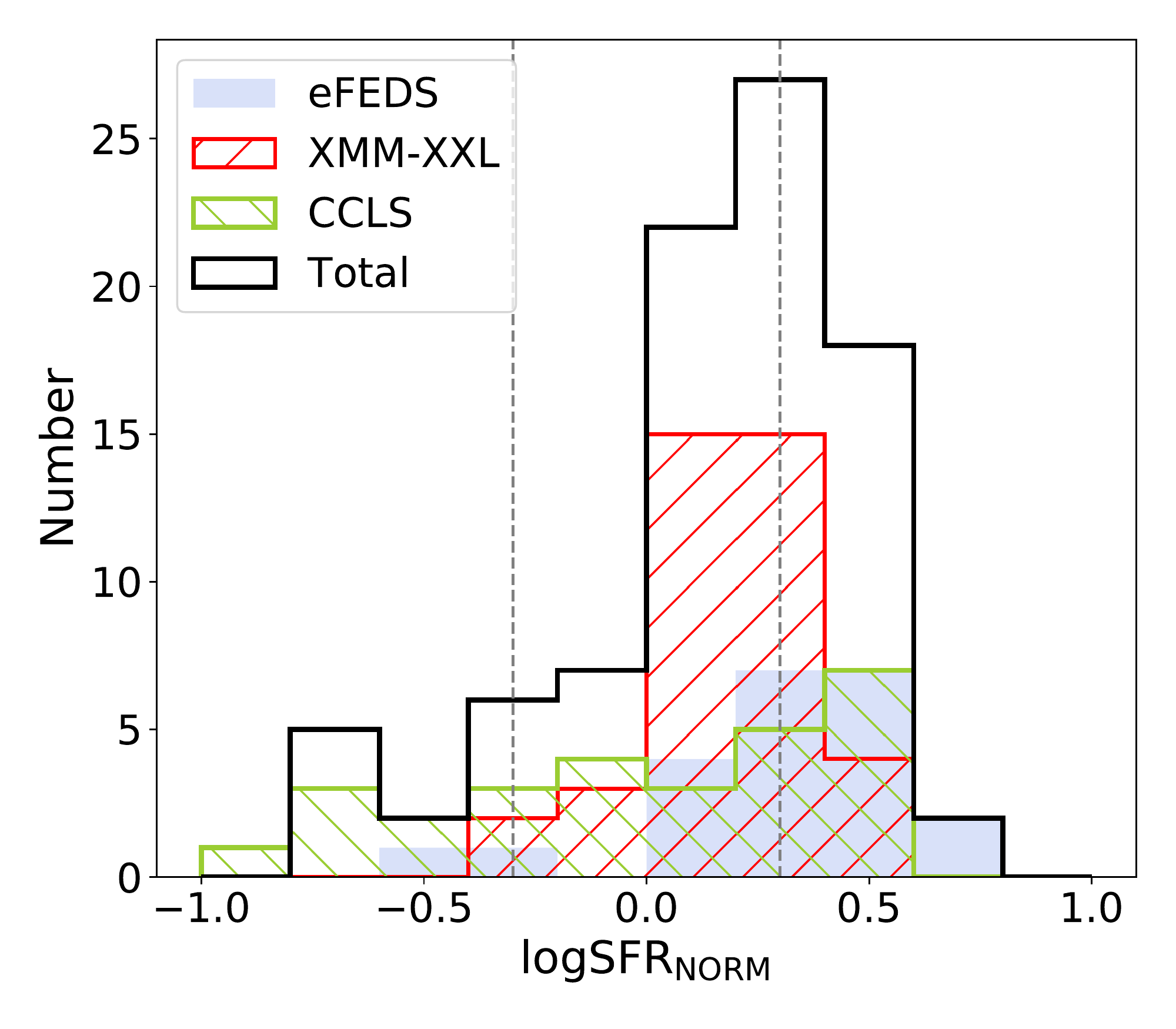} 
    \end{tabular}
\caption{Distributions of $\rm SFR_{NORM}$ for our three samples (as indicated in the legend) and also for the total population (black histogram). The mean and median values of the latter are $\rm log(SFR_{NORM})=0.17$ and $\rm log(SFR_{NORM})=0.20$, respectively. The units are given in logarithmic scale. The vertical dashed lines indicate the limits considered in our analysis for the main sequence of star-forming galaxies ($\pm0.3$ dex).}\label{sfrnorm1}
\end{figure}

A popular way to quantify the position of a source compared to MS consists in calculating the normalised SFR, $\rm SFR_{NORM}$ \citep[e.g.,][]{mullaney2015, masoura2018, bernhard2019, masoura2021}. $\rm SFR_{NORM}$ is the ratio of the SFR of AGNs divided by the SFR of a MS galaxy with similar M$_*$ and redshift as the AGNs. For the latter, we used the analytical expression of \citet{schreiber2015}. In Fig.~\ref{sfrnorm1}, we show the distribution of $\rm SFR_{NORM}$ for the three samples (as indicated in the legend),as well as the total distribution of the ensemble of sources (black histogram). 

Our results show that there is a large fraction of high-z AGNs ($\sim 40\%$) that have enhanced SFR compared to MS galaxies. Taking into account the errors on our SFR measurements (Fig. \ref{m_SFR}), the picture that emerges is that the vast majority of the AGNs in our sample have SFR that is consistent or higher than that of MS star-forming galaxies. In particular, the SFRs of AGNs is increased by a factor of $\sim 1.8$ (mean $\rm log\,SFR_{norm}\sim 0.25$) compared to star-forming galaxies. We note that recent studies \citep{mountrichasBOOTES} have pointed out the caveats of comparing the SFR of X-ray AGNs with that from MS galaxies, using for the latter analytical expressions from the literature \citep[e.g.,][]{schreiber2015}. These systematics are due to the different definitions of the MS, the different selection criteria applied on AGN and non-AGN samples to select sources and the different methods used to estimate host galaxy properties \citep[see Fig. 6 in][]{mountrichasBOOTES}. Even in those cases that the same approach has been followed to calculate the properties of the sources (e.g., SED fitting) the utilization of different templates and different parameter space may affect the measurements (see also next section).

We furthermore noticed that X-ray AGNs in the eFEDS field present the highest fraction of sources above the MS, among the three fields. This is also confirmed by their $\rm SFR_{NORM}$ distribution. eFEDS sources have the highest X-ray luminosities compared to these in COSMOS and XMM-XXL (Fig. \ref{fig_lx_distrib_3fields}). Specifically, the median $\rm L_X$ of AGNs in eFEDS is $\rm L_{2-10~keV}= 10^{45.23}\,erg~s^{-1}$ compared to $\rm L_{2-10~keV}= 10^{44.77}\,ergs~^{-1}$ and $\rm L_{2-10~keV}= 10^{44.34}\,erg~s^{-1}$ of the sources in the XMM-XXL and COSMOS fields, respectively.

\subsection{The role of X-ray luminosity in the position of the AGNs relative to MS}
\label{sec_sfrnorm_lx}
To examine the positions of our X-ray AGNs relative to the MS as functions of the AGN power, we studied the evolution of $\rm SFR_{NORM}$ compared to X-ray luminosity. In our analysis, we used the rest-frame intrinsic (i.e., corrected for absorption, as discussed in Appendix~\ref{xrays}) 2-10 keV luminosities (Fig.~\ref{LX_sfrnorm}). We divided our sample into three $\rm L_X$ bins of equal size ($\sim$30 sources per bin). For each bin, we estimated the median $\rm SFR_{NORM}$ and $\rm L_{2-10 ~ keV}$ values. The uncertainties in each bin were calculated using a bootstrap resampling method. Our results show that $\rm SFR_{NORM}$ is enhanced by $50\%$ within $\sim 1$\,dex in $\rm L_X$ spanned by our sample. Although this increase is not statistical significant ($<1\,\sigma$), $\rm SFR_{NORM}$ values are consistently above the MS (dashed line), which shows that louminous AGNs have enhanced SFR compared to star-forming MS galaxies. In Appendix~\ref{appendixA}, we examined whether the quality criteria applied (Sect.~\ref{criteria}) to our sample could have an impact on these results. Moreover, we tested the case of using only sources with spectroscopic redshifts. These examinations showed that our results are robust.

For comparison, we over-plot data points from previous studies, that have examined the $\rm SFR_{NORM} - L_X$ relation, at lower redshifts. We restricted the comparison to the studies that satisfy the following criteria: $\rm SFR_{NORM}$ has been calculated using the formula of \cite{schreiber2015} and host galaxy properties (SFR, M$_*$) have been measured by applying SED fitting using a similar grid to that used in our study. For all the above cases, the X-CIGALE algorithm has been applied and the same parametric space has been used for the SED fitting, for a fair comparison of the derived properties.

\citet{koutoulidis2021} used X-ray AGNs from the ROSAT-2RXS survey \citep{comparat2020} in the nearby Universe ($\rm z<0.2$) and calculated the $\rm SFR_{NORM}$ parameter for these sources, utilizing the analytical expression of \citet{schreiber2015}. \citet{mountrichasBOOTES} used X-ray AGNs in the Bo$\rm \ddot{o}$tes field within $0.5 < z < 2.0$ and studied the SFR of X-ray AGNs with that of non-AGN systems, by constructing a reference galaxy catalogue. In Fig.~\ref{LX_sfrnorm}, though, we used the values they presented in their Fig. 9, for which they have calculated $\rm SFR_{NORM}$ using the Schreiber et al. formula. \citet{masoura2021} used X-ray AGNs in the XMM-XXL-North field in the redshift range $0.3 < z < 3.5$. In their paper, they calculated SFR and M$_*$ for their X-ray sample using CIGALE. However, their grid is significantly different to that used in our study, since different templates and parameter space have been utilised. To allow a fair comparison, we used their X-ray sample and their photometry for the sources and re-run CIGALE using our grid. We also applied the quality selection criteria described in Sect. \ref{criteria}. 

In the same $\rm L_X$ range, the $\rm SFR_{NORM}$ values calculated for our X-ray sources are in remarkably good agreement with those from previous studies  at lower redshifts. Although the data points from the Bo$\rm \ddot{o}$tes sample appears slightly lower, they are statistically consistent (within 1\,$\sigma$) with our measurements. This indicates that $\rm SFR_{NORM}$ may not evolve with redshift, which confirms the findings of previous studies \citep[e.g.][]{mullaney2015, mountrichasBOOTES} and extends it up to the higher redshifts spanned by our sample. To verify this, we plot the $\rm SFR_{NORM}$ over $\rm L_{X}$ versus redshift (Fig.~\ref{z_sfrnorm}) for our sample along with the aforementioned studies. The Pearson correlation factor between the parameters is $\rm \rho=0.46 \pm 0.12$, suggesting a very weak correlation. The linear fit to the data (using BCES) also supports this low correlation ($y=0.0018 \pm 0.0005 \times x -0.0002 \pm 0.0013$).

We also noticed a difference in the $\rm SFR_{NORM}$ values we calculated using the X-ray sources and the photometry available in the \citet{masoura2021} study using our SED fitting grid compared to the $\rm SFR_{NORM}$ values presented in the  left panel of their Fig. 10. The trends and overall conclusions are the same, i.e., $\rm SFR_{NORM}$ increases with $\rm L_X$, low $\rm L_X$ sources lie below the MS, while high $\rm L_X$ sources are above the MS. However, the $\rm SFR_{NORM}$ values estimated in \citet{masoura2021} are higher by a factor of $\sim 1.5$ on average. This highlights the importance of taking into consideration the different methods and/or grids that are applied in different works to estimate the host galaxy properties when we compare results from different studies. 

Our results corroborate and extend to higher redshift the results from previous studies, that at high $\rm L_X$ ($\rm L_{2-10~keV}>10^{44}$), the SFR of X-ray AGN is enhanced compared to that of star-forming galaxies. This could imply that high luminosity AGN are fuelled by different physical mechanisms than their lower L$_X$ counterparts. For example, galaxy mergers may supply the SMBH with large amounts of cold gas to activate it and at the same time, set off the star formation of the host galaxy. Alternatively, it could be the feedback from the AGN itself that triggers the star formation, by, e.g., over-compressing the cold gas of gas-rich systems, via AGN outflows \citep[e.g.,][]{zubovas2013}.

\begin{figure}
   \begin{tabular}{c}
       \includegraphics[width=0.47\textwidth]{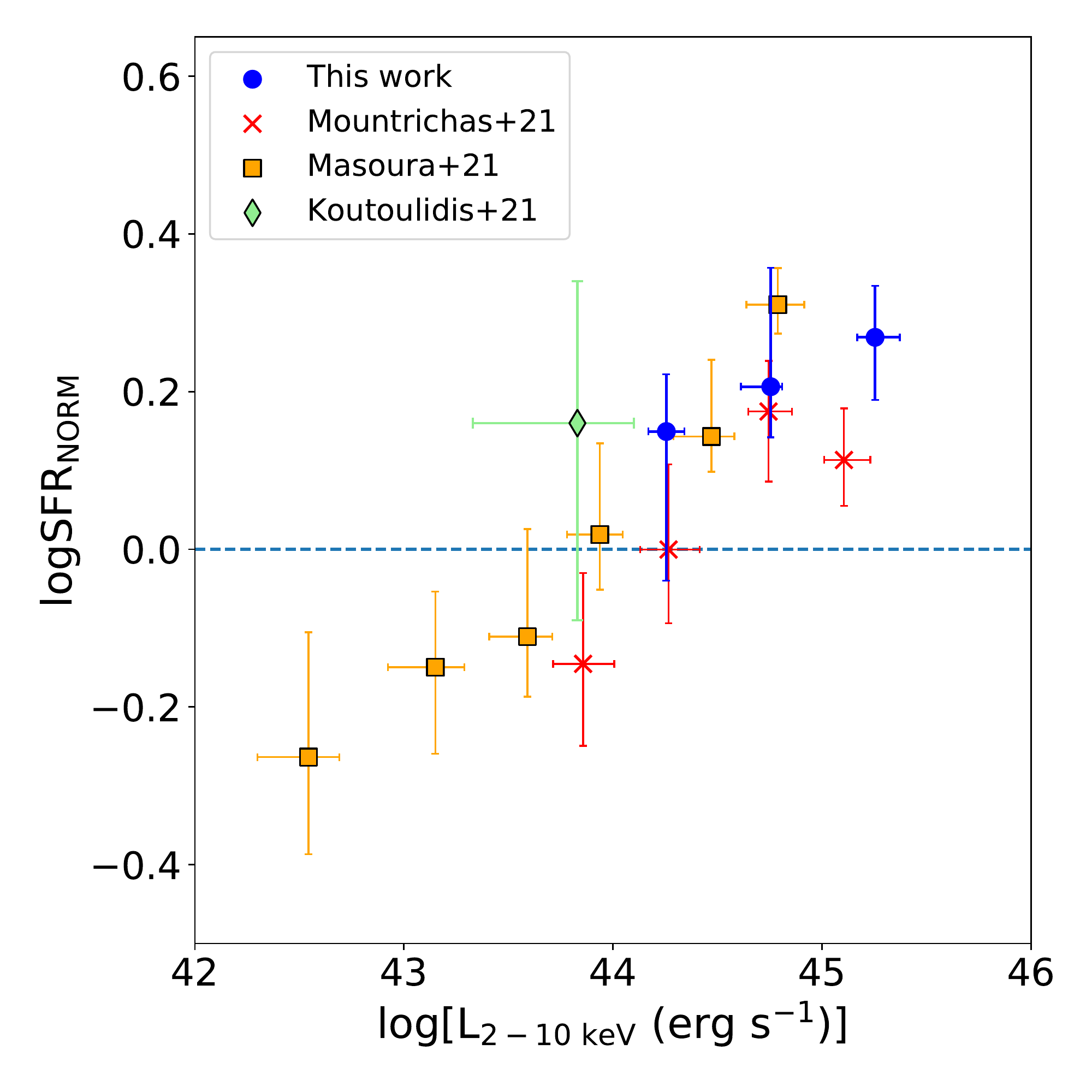} 
    \end{tabular}
\caption{$\rm SFR_{NORM}$ as a function of the rest-frame absorption-corrected X-ray luminosity. Our results are divided into three bins (blue circles) of equal size with uncertainties calculated with bootstrapping. We compare our results with X-ray-selected samples with a variety of luminosity and redshift ranges.}\label{LX_sfrnorm}
\end{figure}

\begin{figure}
   \begin{tabular}{c}
       \includegraphics[width=0.47\textwidth]{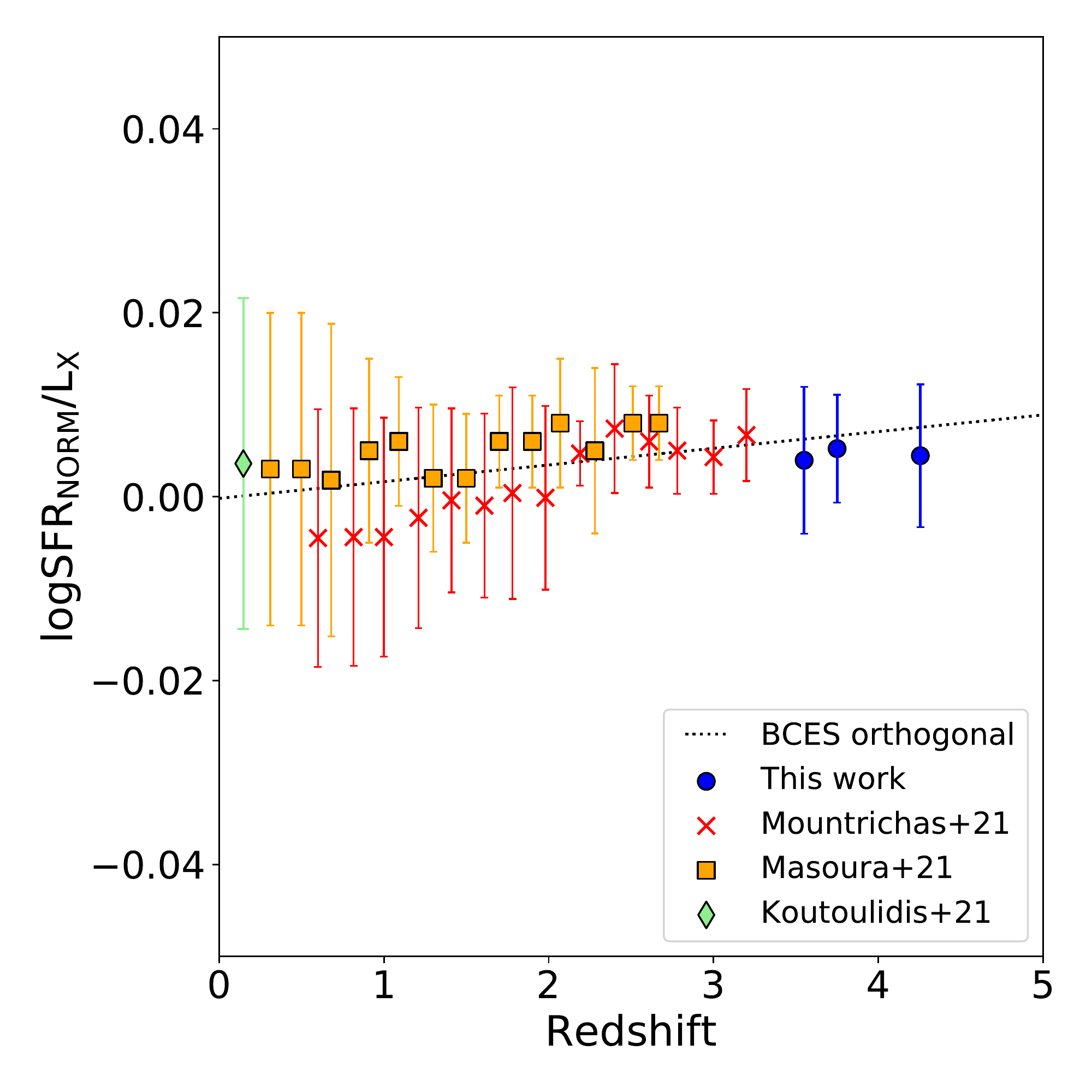} 
    \end{tabular}
\caption{$\rm SFR_{NORM}$ over $\rm L_{X}$ (in logarithmic scale) as a function of redshift. Our results are divided into three bins (blue circles) of equal size with uncertainties calculated with the bootstrap resampling method. We also plot the data points from X-ray-selected samples with a variety of luminosity and redshift ranges as shown in the legend. The dotted line shows the best linear-fit function: $y=0.0018 \pm 0.0005 \times x -0.0002 \pm 0.0013$.}\label{z_sfrnorm}
\end{figure}


\subsection{Eddington ratio and position of the AGN relative to the MS}\label{sectionlambda}

Previous studies found that the M$_*$ of the (host) galaxy affects how the AGN activity and galaxy properties are connected \citep[e.g.,][]{georgakakis2017, aird2018, yang2018, torbaniuk2021}. \cite{mountrichasBOOTES, mountrichas2022} used data in the Bo$\rm \ddot{o}$tes and COSMOS fields and split their sources into luminosity and M$_*$ bins. They found that the SFRs of AGNs are enhanced compared to those of non-AGN systems at $\rm L_{2-10~keV}>10^{44.2}\,erg~s^{-1}$ for systems that have $\rm 10.5<log\,(M_*/M_\odot)<11.5$. However, this may not be true for more massive systems. 

\begin{figure}
      \includegraphics[width=0.48\textwidth]{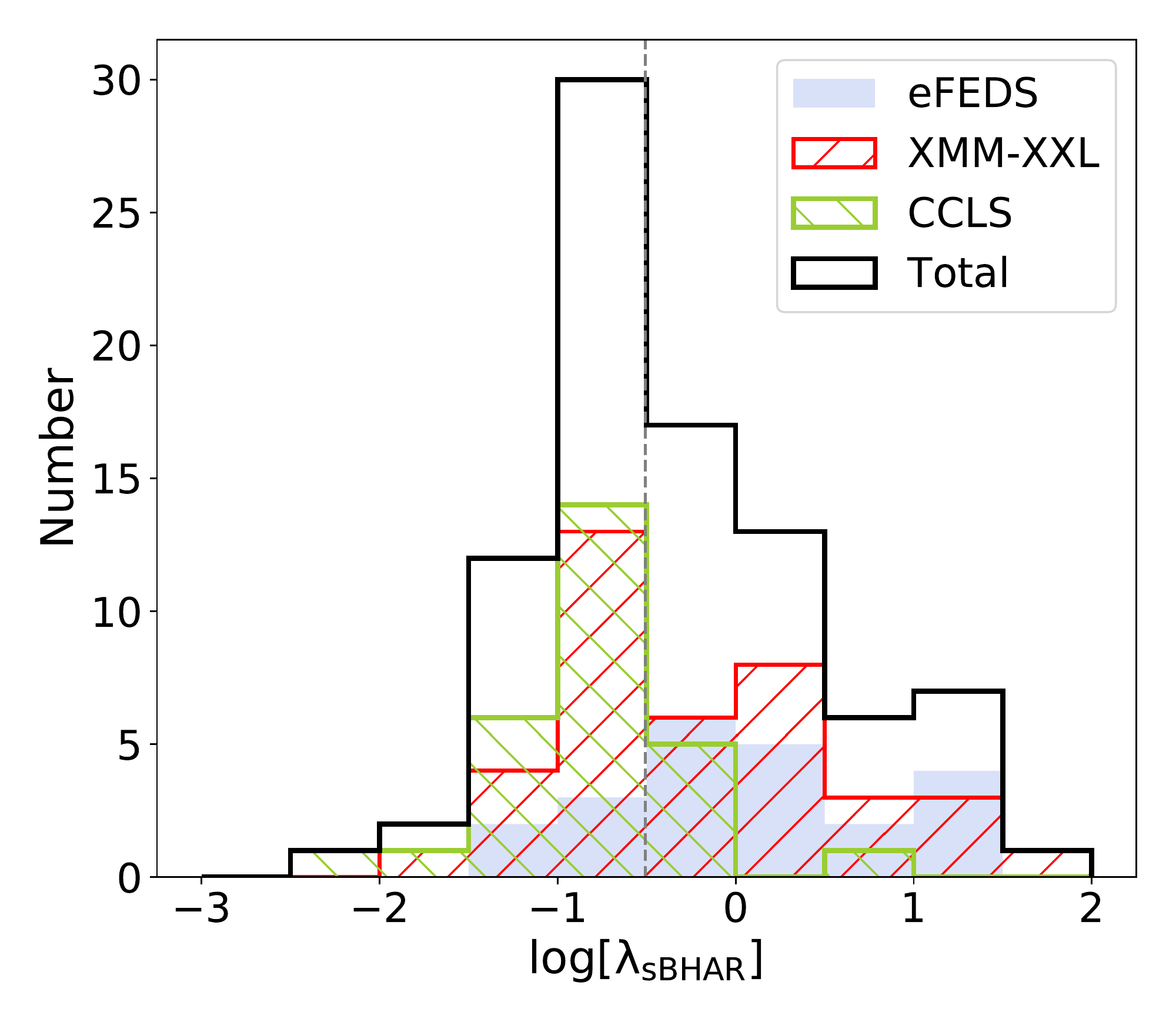}     
\caption{Distribution of $\rm \lambda_{sBHAR}$ (see definition in Sect.~\ref{sectionlambda}) used as a tracer of the Eddington ratio. The median $\rm log \lambda_{sBHAR}$ value of the combined high-z sample (vertical dashed line) is $\sim$-0.5 that corresponds roughly to sources with $\rm \lambda_{Edd}\sim 0.3$.}\label{lambdaHIST}
\end{figure}

The size of our X-ray sample did not allow us to split our measurements in luminosity and M$_*$ bins. To account for the stellar mass of our sources, we calculated the specific black hole accretion rate, $\rm \lambda_{sBHAR}$ \citep[e.g.][]{georgakakis2017}. $\rm \lambda_{sBHAR}$ is defined as the rate of accretion onto the SMBH relative to the M$_*$ of the galaxy. Under the assumption that the BH mass is proportional to M$_*$, $\rm \lambda_{sBHAR}$ can provide a rough measure of the Eddington ratio \citep{Allevato2019}. Following, e.g., \citet{bongiorno2012,bongiorno2016, georgakakis2017, aird2018,aird2019, Allevato2019} this can be written as:

\begin{equation}
\rm \lambda_{Edd}\propto\lambda_{sBHAR}=\frac{k_{Bol}(L_X)\times L_X}{1.3\times10^{38}\rm ergs^{-1}\times A(M_*) \times\frac{M_*}{M_\odot}}, 
\end{equation}
where $\rm k_{Bol}$ is the bolometric correction factor ($\rm L_{Bol}/L_X$), $\rm L_X$ is the X-ray luminosity in the [2-10 keV] band and A is a factor relating the black hole mass with the stellar mass of the host galaxy in units of $\rm M_\odot$. Instead of using constant values of $\rm k_{Bol}$ and A \citep{elvis1994,aird2018}, we calculated the $\rm k_{Bol}$ values that are dependent on the X-ray luminosities \citep{lusso2012, duras2020} using the functional form defined in Eq. 1 of \citet{duras2020} and adopting the coefficients a, b and c from their Table 1. Regarding the black-hole to stellar mass ratio A, we used the Eq. 11 of \citet{kormendy2013} that takes into account the stellar-mass dependence. Figure~\ref{lambdaHIST} presents the $\rm \lambda_{sBHAR}$ distribution of our high-z sample. The median value for the whole sample is $\rm log\lambda_{sBHAR}=-0.51 \pm 0.07$. This corresponds roughly to Eddington ratio values of $\rm \lambda_{Edd}\sim0.3$. Table~\ref{propertiesTable} presents the median values for the individual datasets and their uncertainties derived using a bootstrap resampling method.

\begin{figure}
   \begin{tabular}{c}
     \includegraphics[width=0.5\textwidth]{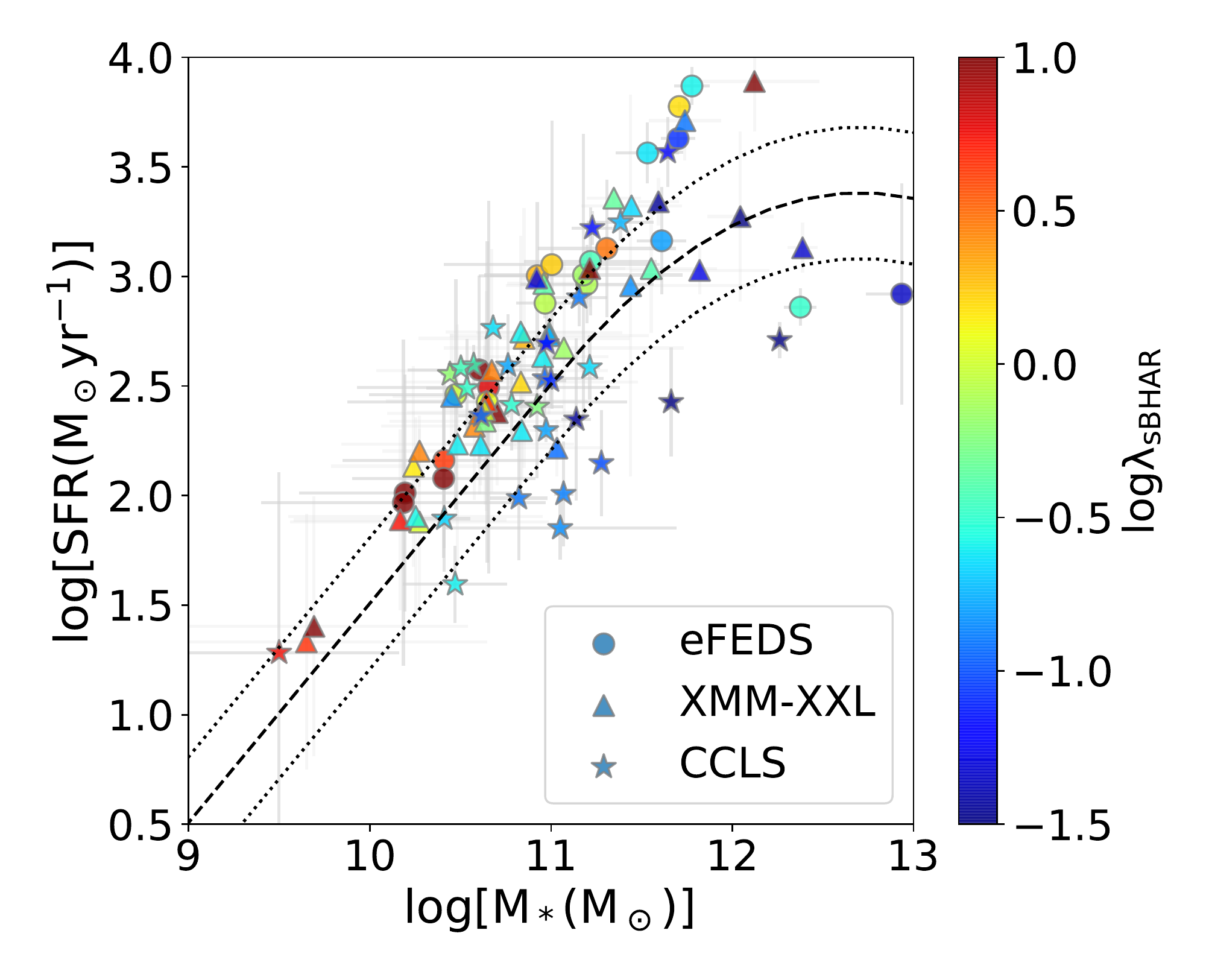} 
    \end{tabular}
\caption{SFR as a function of M*, colour-coded based on $\rm log\lambda_{sBHAR}$ values. The different shapes represent the samples used in our analysis as indicated in the legend.}\label{lambda}
\end{figure}

In Fig.~\ref{lambda}, we present the distribution of the high-z sources in the SFR-M$_*$ plane colour-coded with the corresponding $\rm \lambda_{sBHAR}$. The majority of the sources that lie inside or above the MS of star-forming galaxies, especially at lower M$_*$ ($\rm M_*\leq 10^{11.5} M_{\odot}$), have higher specific accretion rates compared to those of sources below the MS. \citet{setoguchi2021} resulted in similar conclusions using luminous AGNs ($\rm 10^{44.5}<L_{Bol}<10^{46.5}\,erg~s^{-1}$) at $z=1.4$ in the Subaru/XMM-Newton Deep Field (SXDF). In particular, they found that AGNs with high Eddington ratios tend to reside in host galaxies with enhanced normalised star formation. This is also in agreement with the results of \citet{aird2019}. They showed that galaxies with high SFR have higher probability to host an AGN with high specific accretion rate. We notice that at  $\rm logM>10^{12}\,M_{\odot}$ the SFR-M$_*$ relation, as defined by the \cite{schreiber2015} equation, becomes flat. At similar stellar masses, the SFRs of our X-ray AGNs appear constant ($\rm SFR \sim 500-600\,M_\odot yr^{-1}$). This implies a constant  $\rm SFR_{NORM}$ value at this M$_*$ regime, in agreement with the findings of \cite{mountrichasBOOTES, mountrichas2022}. We note, that in Mountrichas et al., in this stellar mass range the SFRs of AGNs appear similar to that of non-AGN systems. However, based on our measurements, the SFRs of AGNs appear lower than those of MS galaxies. Nevertheless, the X-ray luminosities in Mountrichas et al. are at $\rm L_{2-10~keV}<10^{44.6}\,erg~s^{-1}$. In our high-z sample, the vast majority of AGNs, in particular those hosted by such massive galaxies, have $\rm L_{2-10~keV}>10^{45}\,erg~s^{-1}$, denoting higher $\rm \lambda_{sBHAR}$ and, thus, higher Eddington ratios, than their lower redshift counterparts within the same stellar mass range, in the COSMOS and Bo$\rm \ddot{o}$tes fields. This may indicate that higher accreting SMBHs may quench the SFR in the most massive host galaxies. As we have already noted, \citet{mountrichasXXL,mountrichasBOOTES,mountrichas2022} compared the SFR of the X-ray AGNs with that of non-AGN systems by constructing a reference galaxy catalogue instead of using an analytical expression from the literature. Thus, we cannot draw strong conclusions by making a quantitative comparison between our results and theirs. Nevertheless, our results are in qualitative agreement with their findings, that $\rm SFR_{NORM}$ remains roughly constant in the most massive systems.

\section{Summary and conclusions}\label{summary}
In this work, we built a sample of high-redshift AGNs selected in fields with different observed areas and depths (CCLS, XMM-XXL and eFEDS) to examine the relation between the AGN power and the host galaxy properties. We constructed their SEDs using data from X-rays up to FIR and derived the physical properties of their host galaxies using the X-CIGALE SED fitting algorithm. After applying several quality and reliability selection criteria, we obtained an X-ray luminous ($\rm L_{2-10keV} > \sim 10^{44}\,erg\,s^{-1}$) sample of 89 high-z sources ($z\geqslant3.5$) with reliable estimated values of the SFR and the M$_*$ of the host galaxies. 55\% have secure spectroscopic redshifts. Our main results can be summarised as follows:

\begin{enumerate}

\item Using the derived host galaxy properties, we found that our luminous high-z X-ray AGNs live in galaxies with median $\rm M_{*}=5.6 \times10^{10} M_\odot$ and median $\rm SFR_{*}\approx 240\,M_\odot yr^{-1}$. Our results are in agreement with previous studies at high redshifts.

\item We compared the location of our X-ray sources in the SFR-M* plane with that of star-forming MS galaxies adopting the \citet{schreiber2015} relation. We found that $\sim$89\% of our high-z AGNs lie inside (47.2\%) or above the MS (41.6\%), indicating higher star forming activity compared to normal galaxies. In particular, by estimating the $\rm SFR_{NORM}$ parameter, we found that our high-z AGNs have enhanced SFR compared to MS star-forming galaxies by a factor of $\sim 1.8$ (mean $\rm log\, SFR_{NORM}=0.22$), in the luminosity regime probed by our dataset ($\rm L_{2-10~keV} \approx 10^{44-46}\,erg\,s^{-1}$).

\item Our estimated $\rm SFR_{NORM}$ values are in agreement with previous studies at lower redshifts. This is evidence that $\rm SFR_{NORM}$ does not evolve with redshift, confirming the findings of previous studies and extending it up to the higher redshifts probed by our sample.

\item To examine the role of stellar mass in the SFR of our X-ray sample, we used the specific black hole accretion rate, which is a tracer of the Eddington ratio. The bulk of AGNs that lie inside or above the MS of star-forming galaxies have higher specific accretion rates compared to sources below the MS.

\item In agreement with previous studies at lower redshifts \citep{mountrichasBOOTES, mountrichas2022}, our results indicate that in the most massive systems ($\rm log\,(M_{*}/ M_\odot) >10^{11.5-12}$), $\rm SFR_{NORM}$ remains roughly constant.

\end{enumerate}

We have combined data from three different fields and compiled a luminous X-ray sample at very high redshift ($\rm z \geq 3.5$). Although the size of the sample is not large, we have reached important conclusions regarding the SFRs of AGNs in the early Universe and how they compare with those of MS galaxies. Increasing the number of AGNs with available multi-wavelength photometry at high redshifts, will allow us to examine in a statistically significant manner whether $\rm SFR_{NORM}$ increases with $\rm L_X$ and whether our conclusions hold for the most massive systems. Compilation of a large non-AGN galaxy sample at similar redshifts and with similar available photometry will enable us to directly compare the SFRs of AGNs with those of non-AGN systems, minimizing the effect of any systematic effects.

\begin{acknowledgements}\label{ackn}
     The authors are grateful to the anonymous referee for a careful reading and helpful feedback. We acknowledge Lucio Chiappetti for the examination of the manuscript. GM acknowledges support by the Agencia Estatal de Investigación, Unidad de Excelencia María de Maeztu, ref. MDM-2017-0765. IG acknowledge financial support by the European Union's Horizon 2020 programme "XMM2ATHENA" under grant agreement No 101004168. The research leading to these results has received funding (EP and IG) from the European Union's Horizon 2020 Programme under the AHEAD2020 project (grant agreement n. 871158).  
     
     XXL is an international project based around an XMM Very Large Programme surveying two 25 $\rm deg^2$ extragalactic fields at a depth of $\rm \sim 6 \times 10^{-15} erg~s^{-1}~cm^{-2}$ in the [0.5-2] keV band for point like sources. The XXL website is http://irfu.cea.fr/xxl. Multi-band information and spectroscopic follow-up of the X-ray sources are obtained through a number of survey programmes, summarised at \url{http://xxlmultiwave.pbworks.com/}. This research made use of Astropy, a community-developed core Python package for Astronomy \citep[\url{http://www.astropy.org},][]{astropy2018}. This publication made use of TOPCAT \citep{taylor2005} for table manipulations. The plots in this publication were produced using Matplotlib, a Python library for publication quality graphics \citep{hunter2007}.
     
     Based on observations obtained with XMM-Newton, an ESA science mission with instruments and contributions directly funded by ESA member states and NASA. This work is based on data from eROSITA, the soft X-ray instrument aboard SRG, a joint Russian-German science mission supported by the Russian Space Agency (Roskosmos), in the interests of the Russian Academy of Sciences represented by its Space Research Institute (IKI), and the Deutsches Zentrum für Luft- und Raumfahrt (DLR). The SRG spacecraft was built by Lavochkin Association (NPOL) and its subcontractors, and is operated by NPOL with support from the Max Planck Institute for Extraterrestrial Physics (MPE). The development and construction of the eROSITA X-ray instrument was led by MPE, with contributions from the Dr. Karl Remeis Observatory Bamberg \& ECAP (FAU Erlangen-Nuernberg), the University of Hamburg Observatory, the Leibniz Institute for Astrophysics Potsdam (AIP), and the Institute for Astronomy and Astrophysics of the University of Tübingen, with the support of DLR and the Max Planck Society. The Argelander Institute for Astronomy of the University of Bonn and the Ludwig Maximilians Universität Munich also participated in the science preparation for eROSITA. This research has made use of data obtained from the Chandra Data Archive and the Chandra Source Catalog, and software provided by the Chandra X-ray Center (CXC) in the application packages CIAO and Sherpa.

    The Hyper Suprime-Cam (HSC) collaboration includes the astronomical communities of Japan and Taiwan, and Princeton University. The HSC instrumentation and software were developed by the National Astronomical Observatory of Japan (NAOJ), the Kavli Institute for the Physics and Mathematics of the Universe (Kavli IPMU), the University of Tokyo, the High Energy Accelerator Research Organization (KEK), the Academia Sinica Institute for Astronomy and Astrophysics in Taiwan (ASIAA), and Princeton University. Funding was contributed by the FIRST program from the Japanese Cabinet Office, the Ministry of Education, Culture, Sports, Science and Technology (MEXT), the Japan Society for the Promotion of Science (JSPS), Japan Science and Technology Agency (JST), the Toray Science Foundation, NAOJ, Kavli IPMU, KEK, ASIAA, and Princeton University. This paper makes use of software developed for the Large Synoptic Survey Telescope. We thank the LSST Project for making their code available as free software at  http://dm.lsst.org. This paper is based [in part] on data collected at the Subaru Telescope and retrieved from the HSC data archive system, which is operated by Subaru Telescope and Astronomy Data Center (ADC) at National Astronomical Observatory of Japan. Data analysis was in part carried out with the cooperation of Center for Computational Astrophysics (CfCA), National Astronomical Observatory of Japan.

\end{acknowledgements}

%

\bibliographystyle{aa}

\bibliography{aanda} 
%




\begin{appendix}

\section{X-ray properties}\label{xrays}

X-CIGALE has the ability to include the X-ray flux, $\rm f_X$, in the SED fitting process. For that, it requires the intrinsic, i.e., the absorption corrected, $\rm f_X$. Thus, in our analysis we used for all the X-ray AGNs in the three fields the rest-frame intrinsic (unabsorbed) X-ray luminosity in the $2-10$\,keV hard X-ray band. Regarding the CCLS sample, \citet{marchesi2016}, provides a luminosity absorption correction factor, obtained through hydrogen column density, $\rm N_H$, measurements. $\rm N_H$ quantifies the X-ray absorption of each source. We used these correction estimates to convert the observed X-ray luminosities into the intrinsic luminosities, required by X-CIGALE. The $\rm N_H$ calculations were derived from hardness ratio, HR, estimates (defined as the ratio $\rm \frac{H-S}{H+S}$, where H and S are the net counts of the sources in the hard and in the soft band, respectively). However, for sources with more than 70 counts, $\rm N_H$ has been estimated following an X-ray spectral fitting analysis \citep{marchesi2016spectra}. Four such sources are included in our high-z sample. For them, we corrected their luminosity based on the $\rm N_H$ value, estimated via spectral fitting. We confirmed that the two estimates are in good agreement and, in any case, this choice does not affect our SED fitting results. In eFEDS, we used the X-ray luminosities provided by \citet{liu2021} obtained using spectral analysis. We used the results of Model I of the AGN emission that includes a single power-law model (TBabs*zTBabs*powerlaw in Xspec terminology) allowing to constrain the photon index, $\Gamma$, and the hydrogen column density. \citet{liu2021} used a Gaussian prior for $\Gamma$ with values between -2 and 6, and a log-uniform prior for the estimation of $\rm N_H$ between $4\times10^{19}$ and $4\times10^{24}$ $\rm cm^{-2}$

To calculate the rest-frame corrected X-ray luminosities of the high-z AGNs in the XMM-XXL field, we analysed the X-ray spectra as described below. X-ray spectra were extracted using SAS 19.0.0, following the standard procedure outlined by the SAS documentation. Source spectra were extracted in circular regions using a 30 arcsec radius, centred at the position of the X-ray source given by the catalogue. The corresponding background spectra for each source were extracted in circular regions of 30 arcsec, centred in a position of 1.5 arcmin from the source. The exact position of the background region was selected to maximise the number of good pixels in the region (after masking areas outside the detectors and other nearby detected sources) and to be as close as possible to the detector column of the source. We used the pre-calculated redistribution matrices, selecting the proper ones depending on the detector position and epoch of observation. 

\begin{figure}
   \begin{tabular}{c}
       \includegraphics[width=0.47\textwidth]{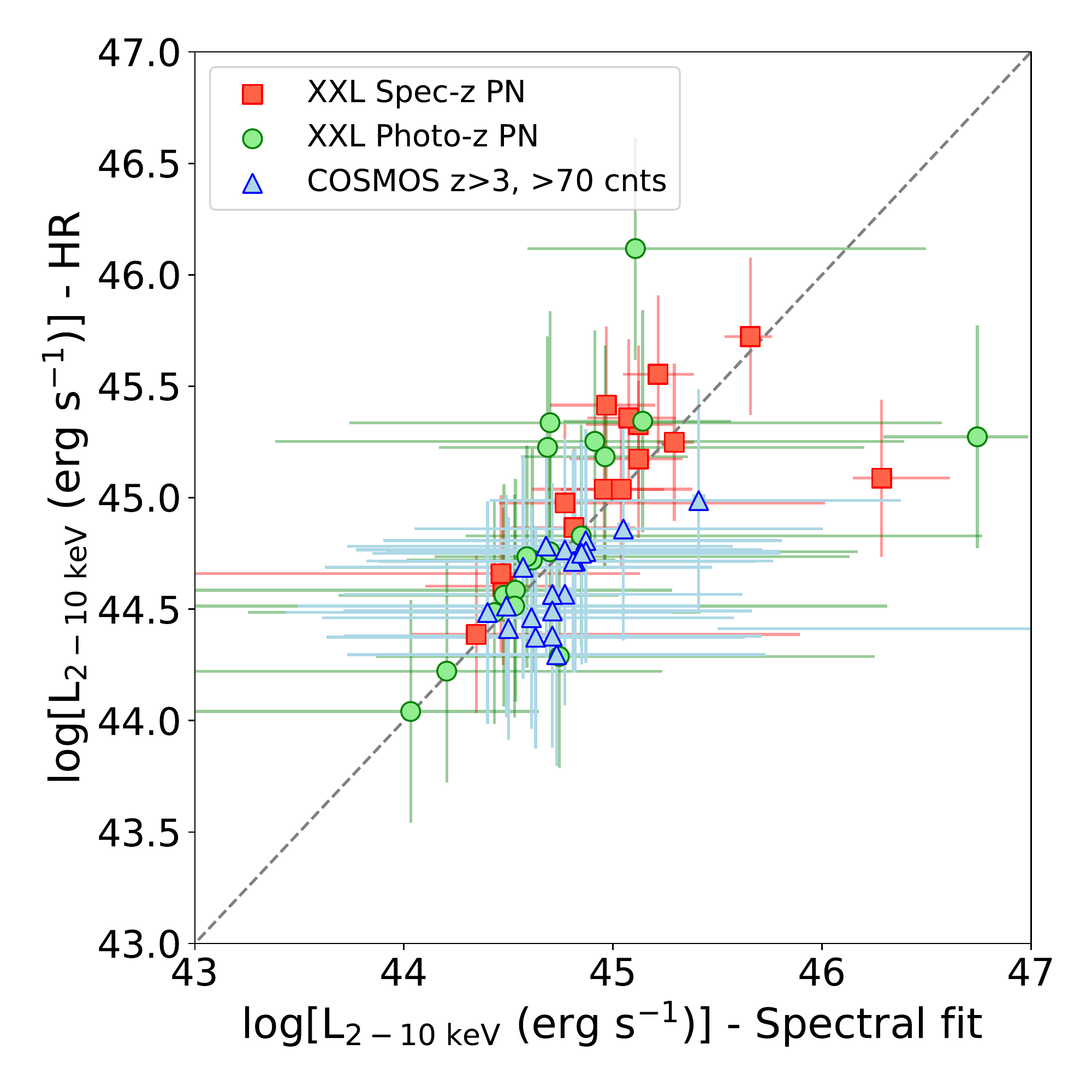} 
    \end{tabular}
\caption{Comparison of the X-ray absorption-corrected luminosity derived with hardness ratio and spectral fitting. The triangles correspond to the 20 sources with z>3 and net counts larger than 70 in the COSMOS field. The squares (circles) represent the XXL sources with spectroscopic (photometric) redshift detected with the PN camera.}\label{hr_spectral}
\end{figure}

For analysing X-ray spectra, we used the analysis software BXA \citep[version 4.0.2,][]{buchner2014}, which connects the nested sampling algorithm UltraNest \citep{buchner2021} with the fitting environment CIAO/Sherpa \citep[version 4.13,][]{freeman2001,fruscione2006}. All spectra and observations obtained by all the EPIC camera data for a given source were simultaneously fitted at once using the UXClumpy torus model with a scattering component \citep{buchner2019} plus a multiplicative absorption component to take into account the Galactic Hydrogen column density (tbabs). Our model have five free parameters: the $\rm N_H$ of the torus, the Photon index of the unabsorbed X-ray emission ($\Gamma$), the inclination angle of the torus with respect to the line-of-sight of the observer and the normalisations for the torus and scattering components. We used Jeffreys priors (flat in log scale) for the $\rm N_H$ (with limits between 20-26), the inclination angle (0-90 degrees) and the normalisations. For the photon, index we assumed a Gaussian prior with mean 1.9 and standard deviation 0.15 \citep{nandra1994}. In the case of sources with photometric redshifts, we also treated the redshift as a free parameter, using as a prior the probability density function estimated by X-CIGALE in \citet{pouliasis2021}.

The use of hardness ratios in the COSMOS field could introduces some biases to our results. In order to check the reliability of deriving $\rm N_H$ values (hence, absorption-corrected luminosities) from hardness ratios, we used the intrinsic luminosities obtained in \citep{marchesi2016spectra}. However, in this study, they provide only the results for $z>3$ sources with >70 net counts in their spectra. In Fig.~\ref{hr_spectral}, we compare the intrinsic luminosity derived with the two methods for the 20 sources with z>3 and net counts>70 (blue triangles). In addition, we have derived the intrinsic luminosities using the hardness ratio in the XMM-XXL field and plot the results. For the latter, we used the PN data for the photo-z and zspec samples. In general, there is a good agreement between the two methods (hardness ratio, spectral fits) in the computation of the absorption-corrected luminosity. Thus, we do not expect the use of hardness ratios in the COSMOS field to affect significantly our main results.

\section{Reliability of host galaxy properties measurements}\label{appendixB}

\begin{figure}
\center
   \begin{tabular}{c c}
   \includegraphics[width=0.47\textwidth]{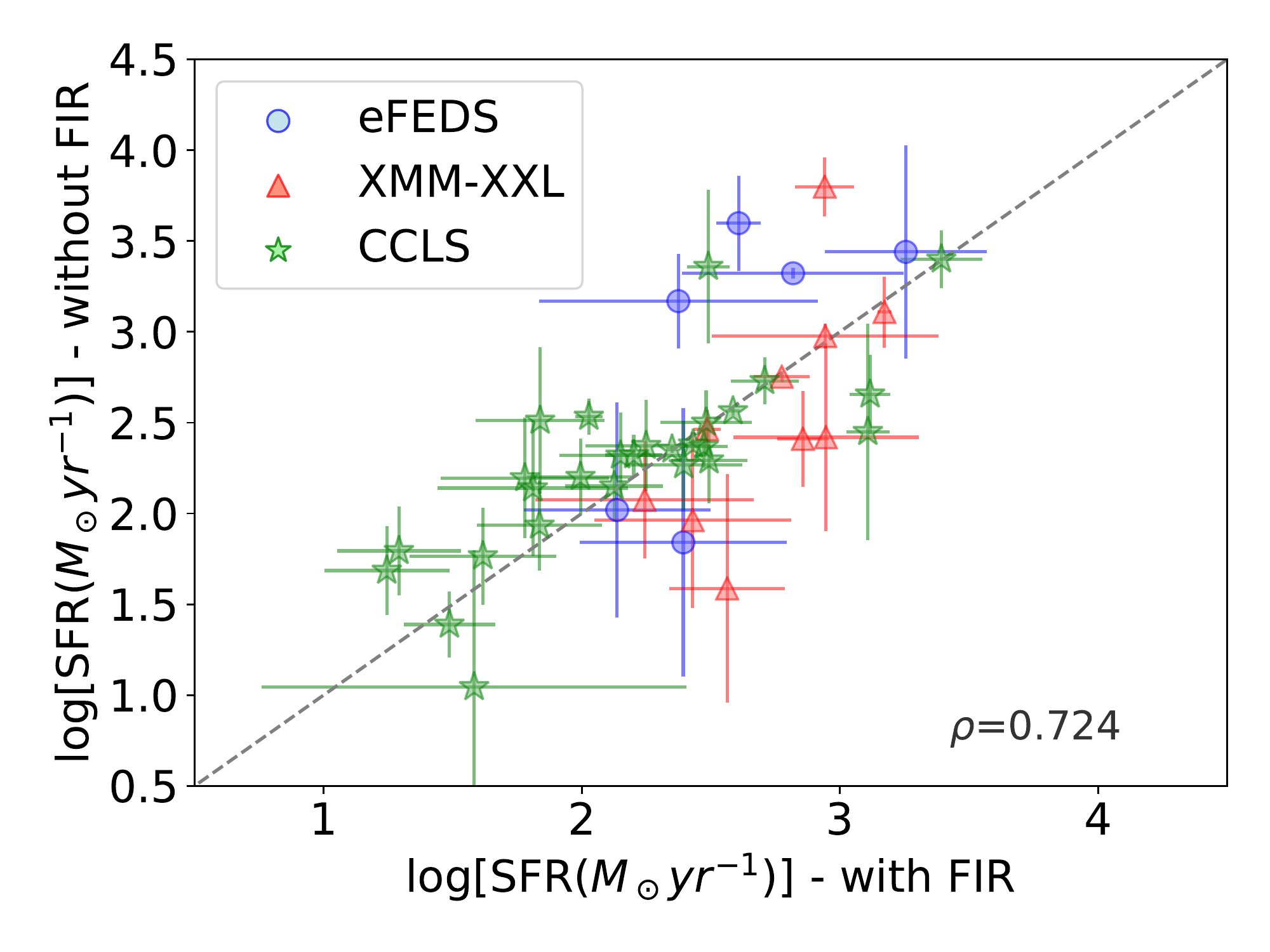} \\
    \includegraphics[width=0.47\textwidth]{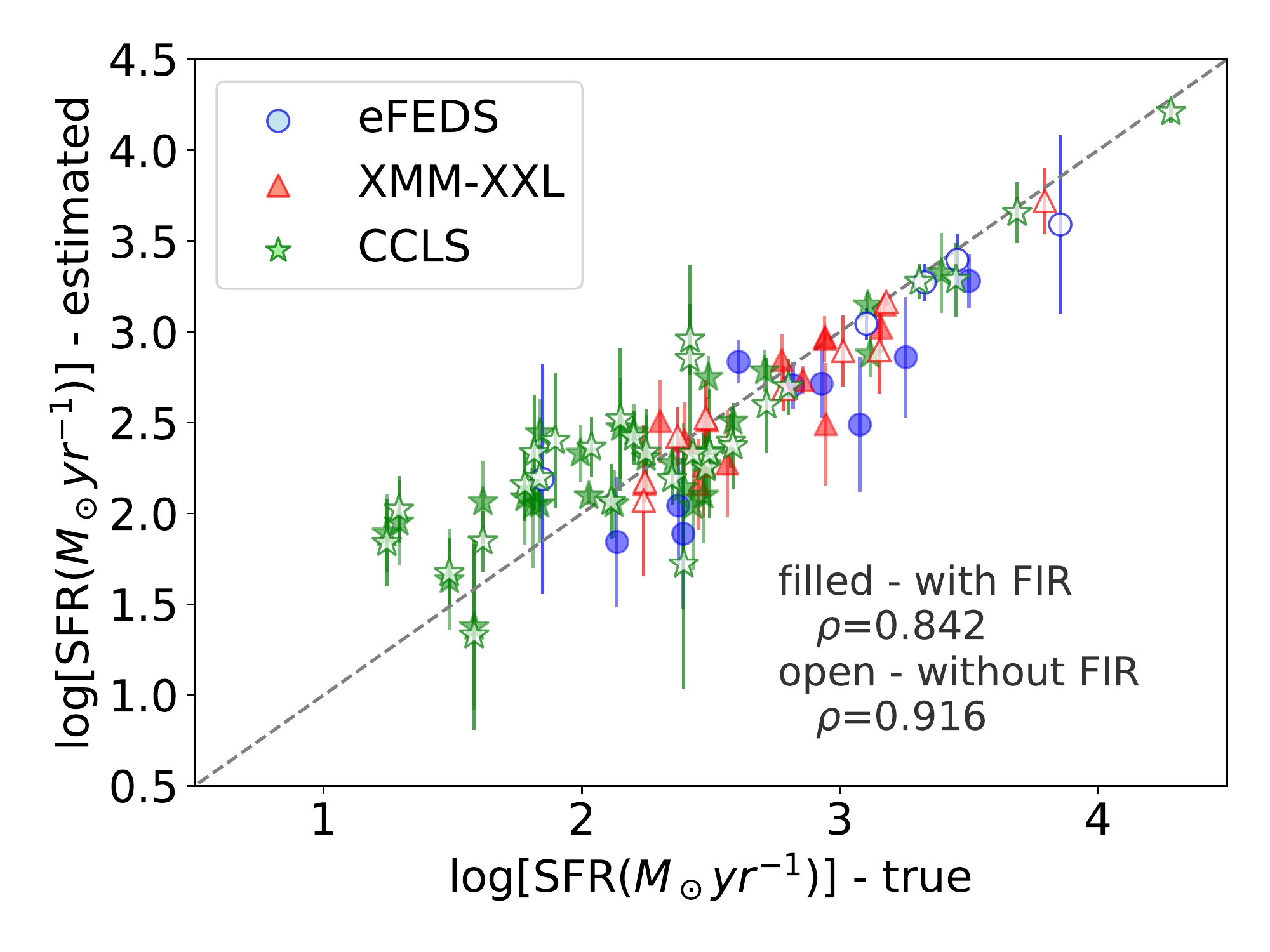} \\
    \includegraphics[width=0.47\textwidth]{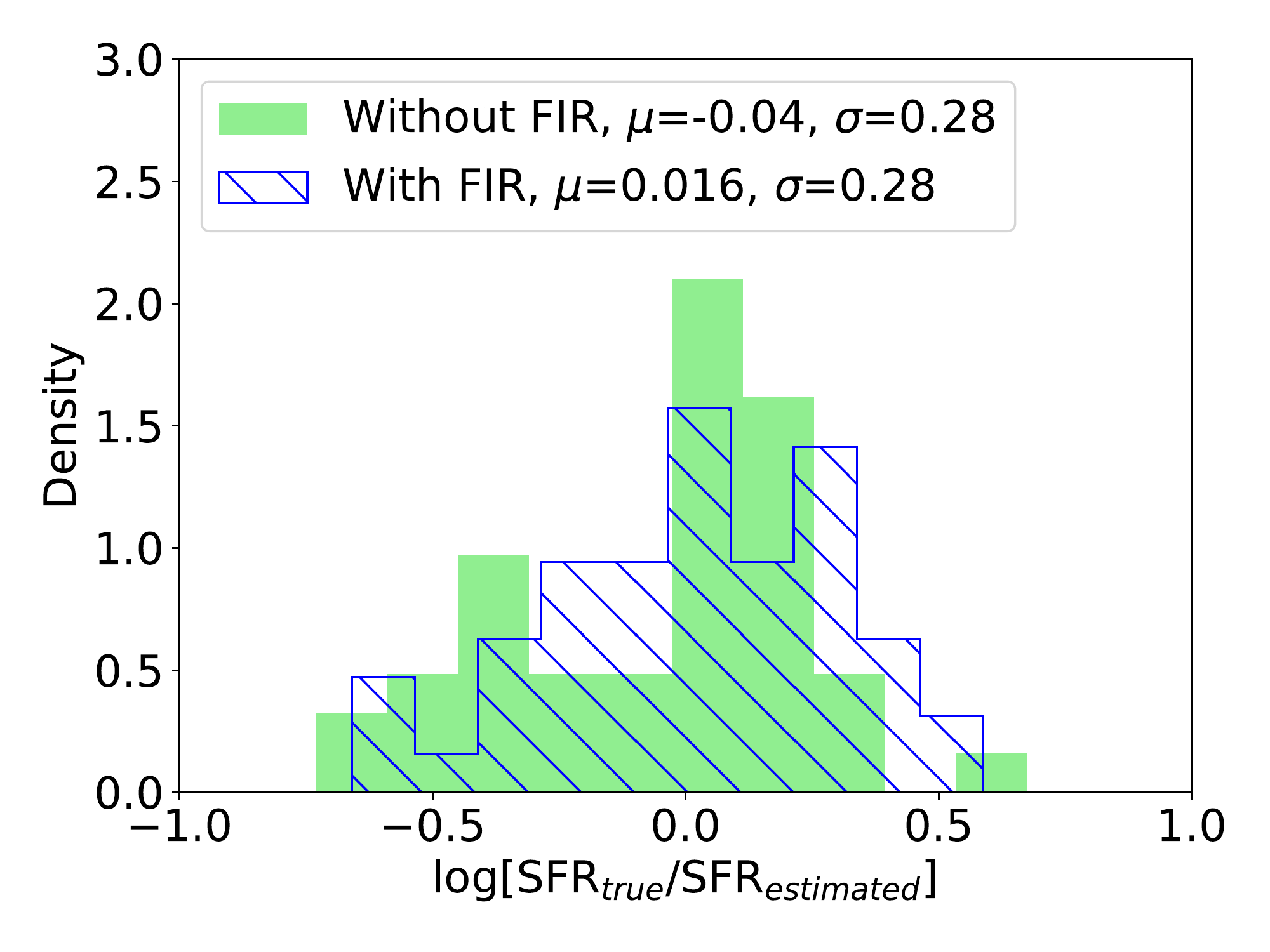}\\
    \end{tabular}
\caption{Top panel: Comparison of SFR  measurements with and without FIR photometry, for the high-z sources that have FIR coverage and satisfy our quality selection criteria (Sect.~\ref{criteria}). The dashed lines lines present the 1:1 relation. The parameter calculations are in good agreement based on the Pearson correlation factor, $\rho$ given in the plot. Middle panel: The estimated SFR based on the mock catalogues compared to the true values provided by the best-fit model for the three fields used in our analysis. The filled (open) symbols represent the X-CIGALE runs with (without) FIR data. The dashed line represent the 1-to-1 relation to assist the plot interpretation, while in each plot we give the linear Pearson correlation coefficient ($\rm \rho$) of the combined sample. Bottom panel: The corresponding distributions of the difference between estimated and true values. We provide also the mean and standard deviation for each case.}\label{mock_results_FIR}
\end{figure}

\begin{figure}
   \begin{tabular}{c c}
      \includegraphics[width=0.47\textwidth]{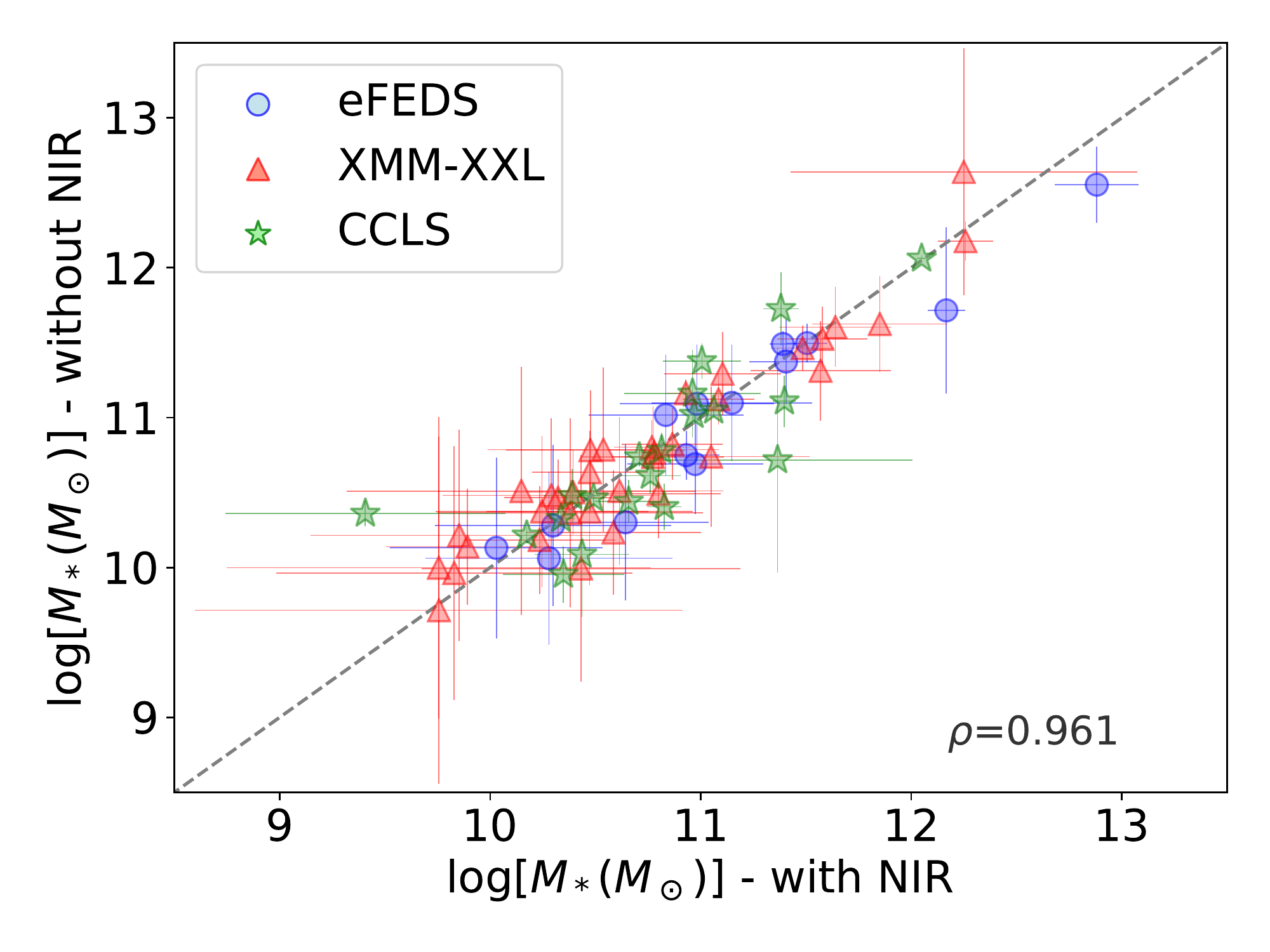} \\
       \includegraphics[width=0.47\textwidth]{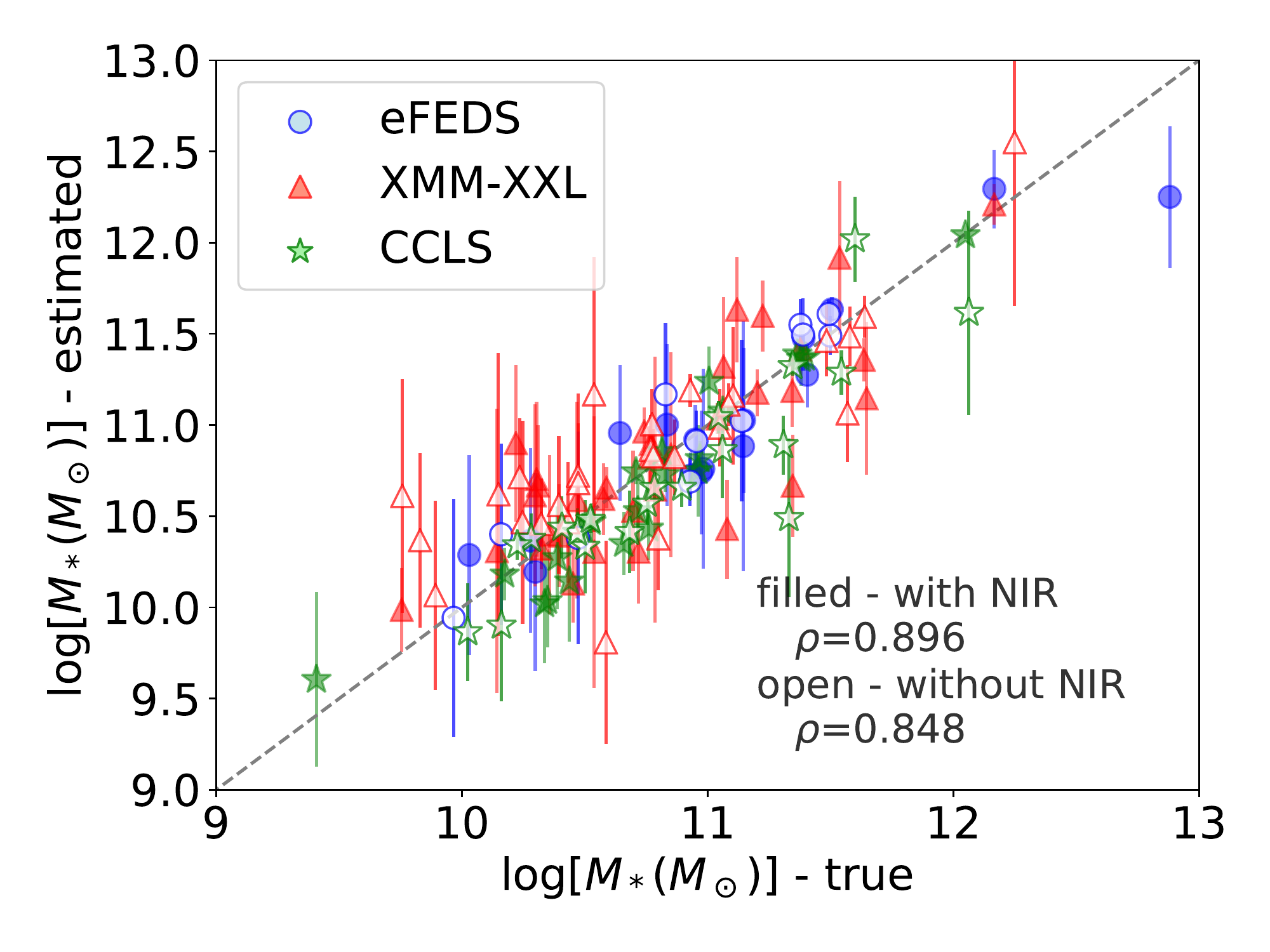} \\
    \includegraphics[width=0.47\textwidth]{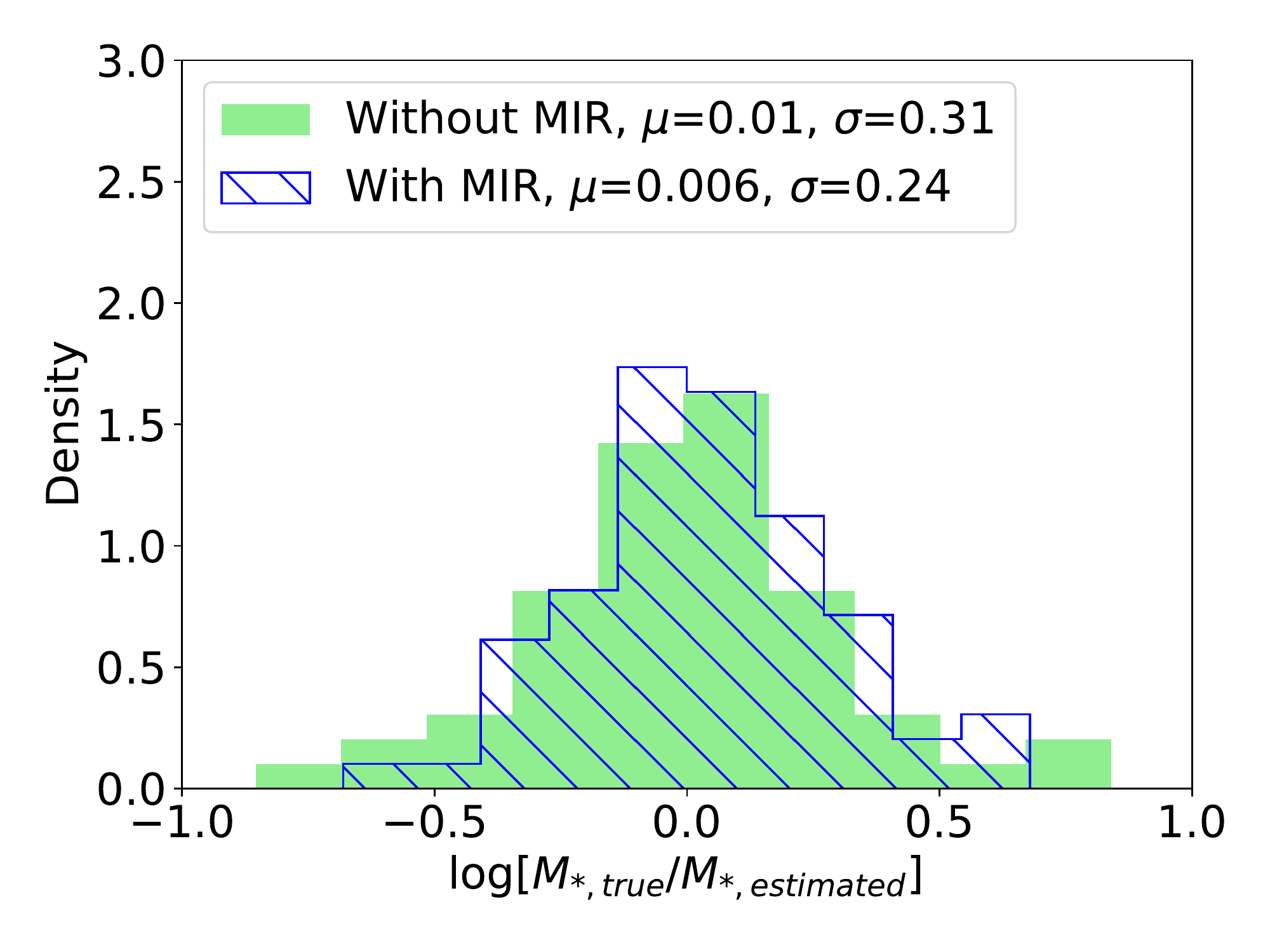}\\
    \end{tabular}
\caption{Top panel: Comparison of stellar mass measurements with and without NIR photometry, for the high-z sources that have NIR coverage and satisfy our quality selection criteria (Sect.~\ref{criteria}). The dashed lines lines present the 1:1 relation. The parameter estimations are in very good agreement considering the Pearson correlation factor, $\rho$, given in the plots. Middle panel: The estimated stellar mass based on the mock catalogues compared to the true values provided by the best-fit model for the three fields used in our analysis. The filled (open) symbols represent the X-XIGALE runs with (without) NIR data. Bottom panel: The distributions of the difference between estimated and true values. The mean difference and its dispersion are very similar in the two runs (with and without NIR photometry).}\label{mock_results_NIR}
\end{figure}

In this section, we examined whether lack of some photometric bands for a (small) number of AGNs in our sample, affected the SFR and M$_*$ measurements and the ability of X-CIGALE to constrain the AGN component. M$_*$ is calculated based on the optical and NIR photometry. Taking into account the high-z selection, W1 or IRAC1 and W2 or IRAC2 photometric bands map he rest-frame NIR part of the spectrum and thus also help to constrain the M$_*$ parameter. FIR along with optical photometry is used by X-CIGALE to estimate the SFR parameter. W1 or IRAC1 and W2 or IRAC2 are used at lower redshifts to constrain the AGN component (torus). However, at the redshift range of our dataset, the longer wavelength data (W4 or MIPS1) are essential to constrain the AGN properties properly.

Out of the final 89 high-z sources, 50 sources (56\%) have available FIR photometry (Herschel PACS and/or SPIRE photometric bands). For these sources, we run X-CIGALE twice. One time including the FIR photometry and a second time without {\it{Herschel}} information, using both times the same parametric grid (Table \ref{proposal}). Our goal was to examine whether lack of FIR data affects the estimated SFR. First, in the top panel of Fig.~\ref{mock_results_FIR}, we compared the SFR measurements of the two runs (with and without FIR data). In both cases, we found a very good agreement between the measurements. Calculating the Pearson correlation factor, we found $\rm \rho_{SFR}=0.724$, indicating good correlation between the two runs. We followed the same procedure to examine whether lack of MIR or NIR affects the SFR calculations and found no significant differences between the results.

We also make use of the mock catalogues that X-CIGALE can produce. In the middle panel of Fig.~\ref{mock_results_FIR}, we show the estimated parameter (mock analysis) of the SFR with and without FIR data compared to the true values, for those sources that satisfied our selection criteria described in Sect.~\ref{criteria}. The corresponding linear Pearson correlation coefficients of the combined samples are $\rm \rho_{SFR}=0.842$ when FIR data are included in the SED fitting analysis, and $\rm \rho_{SFR}=0.916$ when we exclude FIR photometry. On the bottom panel of Fig.~\ref{mock_results_FIR}, we show the distributions of the difference between the estimated and true values of the physical properties for both runs. $\mu$ represents the mean difference with its standard deviation, $\sigma$. When we used FIR photometry, the mean values of the differences are slightly closer to zero while having similar scatter. In both cases, the estimated parameter is well constrained.

In the XXL field, all 39 X-ray sources have available optical photometry ($g, r, i, z$) and 37 out of 39 have detection in at least one NIR band. 32 out of 39 sources have available MIR photometry (W1 or IRAC1 and W2 or IRAC2). The two sources without NIR photometry have available MIR bands. In COSMOS, 28/28 have optical photometry (at least $r, i, z$), 22 out of 28 have NIR, while all  28 sources have {\it{Spitzer}} detection. All 22 sources in eFEDS have been detected in the optical, 16 out of 22 have NIR photometry and are all detected by WISE (W1 and W2). Overall 71 out 89 have optical, NIR and MIR (W1 or IRAC1 and W2 or IRAC2) photometry available. 76 out of 89 have NIR photometry and 82 out of 89 have MIR photometry. There are no sources that lack both NIR and MIR photometry. Next, we examine how lack of NIR bands affects the M$_*$ calculations.

Following the process described above, for those sources that have NIR photometry (76 AGNs), we run X-CIGALE twice. One time taking into account the NIR bands in the SED fitting process and the second time ignoring the NIR information. The comparison of the estimated M$_*$ for the two runs is shown in the top panel of Fig. \ref{mock_results_NIR}. We also use the mock catalogues to examine how well can X-CIGALE constrain M$_*$ in the absence of NIR photometry. The results are shown in middle and bottom panels of Fig. \ref{mock_results_NIR}. Based on these tests, X-CIGALE has measured reliably the M$_*$ for the 13 sources in our dataset that lack NIR observations.

We also examined if X-CIGALE is able to constrain the AGN component without the longest MIR bands (W4 or MIPS1). There are 26 AGNs in our dataset (20 in XXL and 6 in eFEDS) that lack detection in W4 or MIPS. \citet{florez2020} examined if the absence of MIR photometry has an impact on their derived properties and found no significant deviations (their Fig. A.2 and A.3). In our case, the inclusion of the X-ray flux in the SED fitting process should provide X-CIGALE an (additional) constrain for the AGN emission. To confirm the above hypothesis, we run X-CIGALE for the 63 AGNs that have W4 or MIPS band available, using the same parametric grid, and excluding these bands from the fitting process.

\begin{figure}
   \begin{tabular}{c}
       \includegraphics[width=0.47\textwidth]{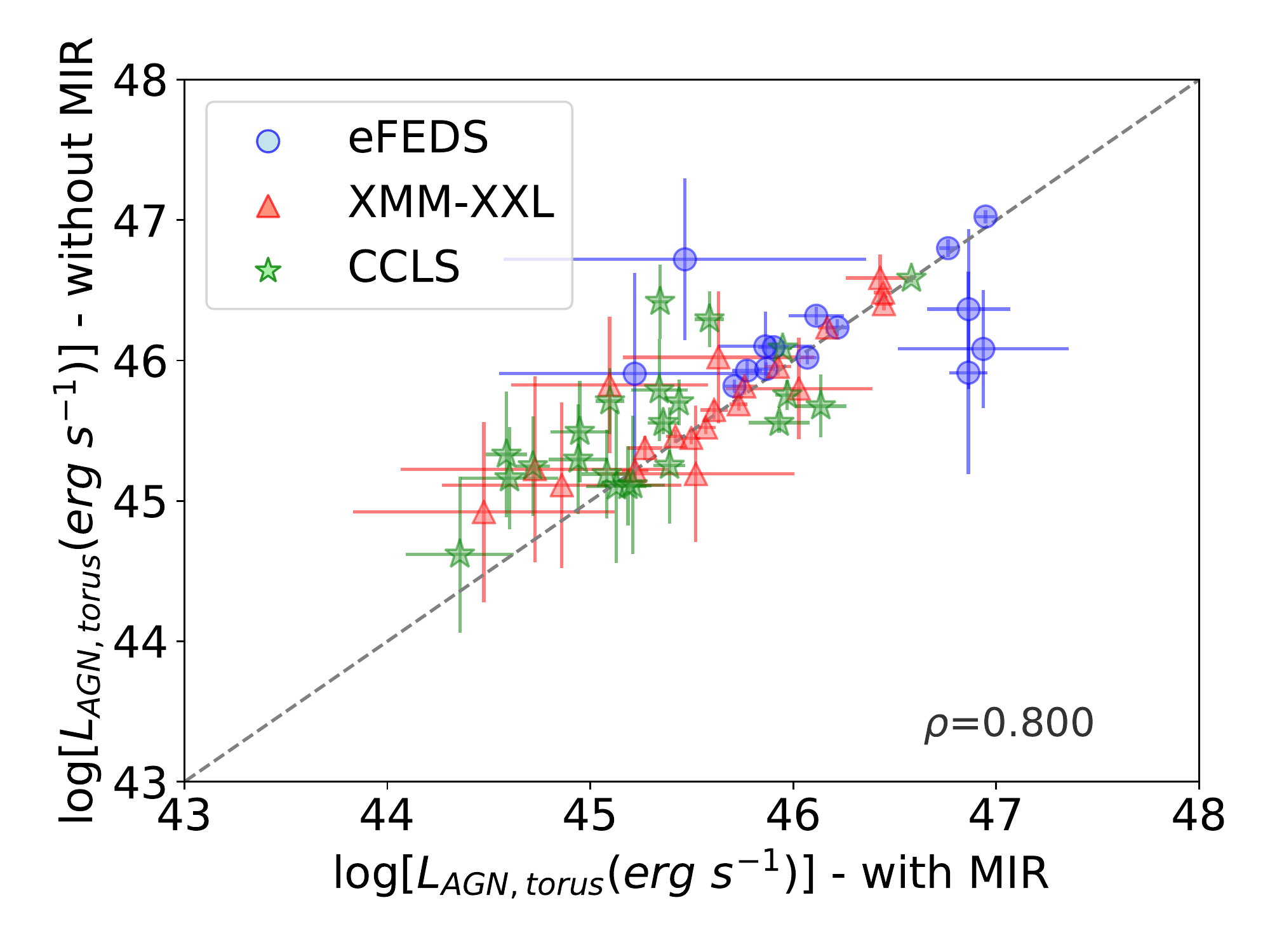} \\
        \includegraphics[width=0.46\textwidth]{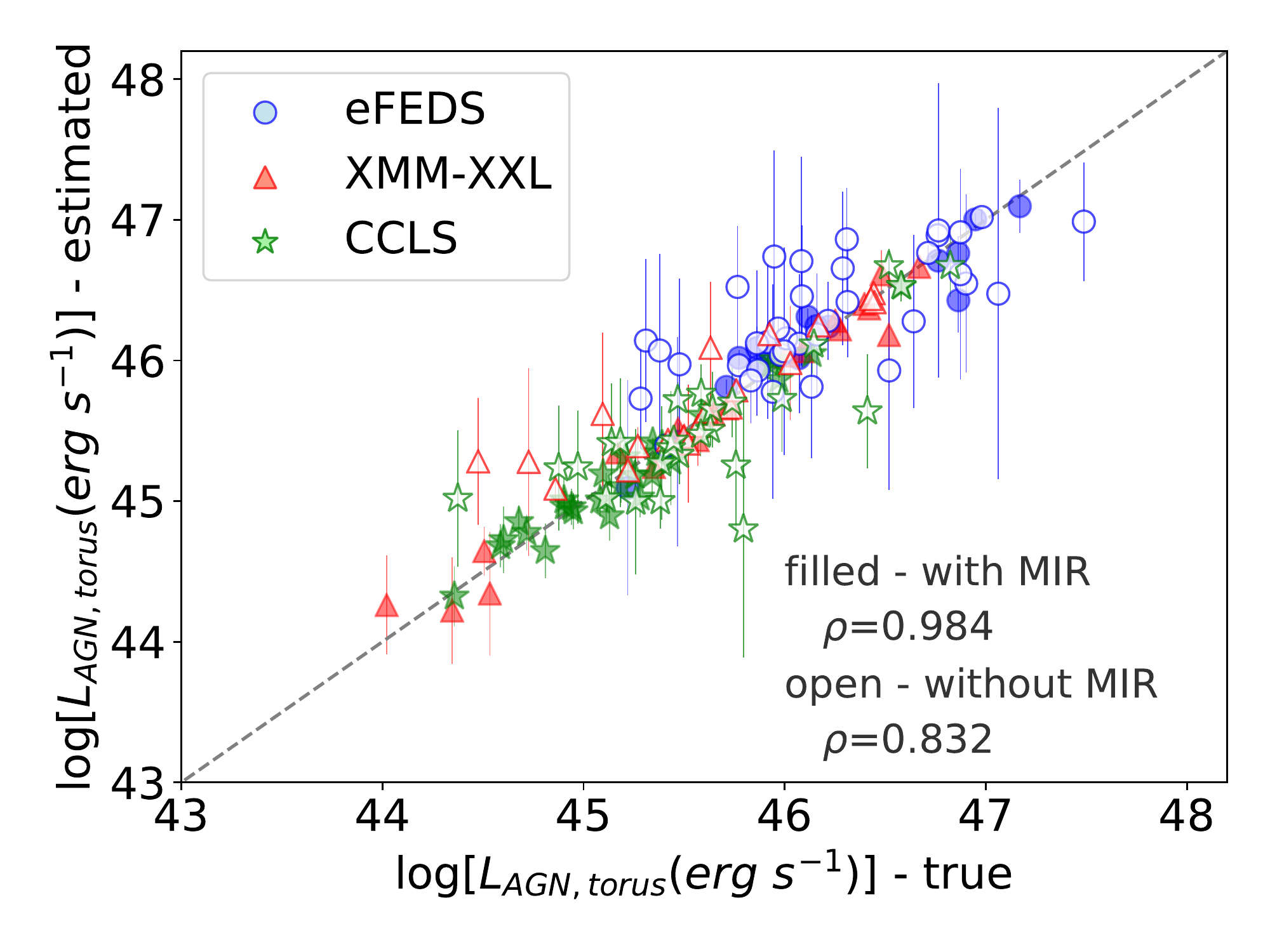} \\
    \includegraphics[width=0.48\textwidth]{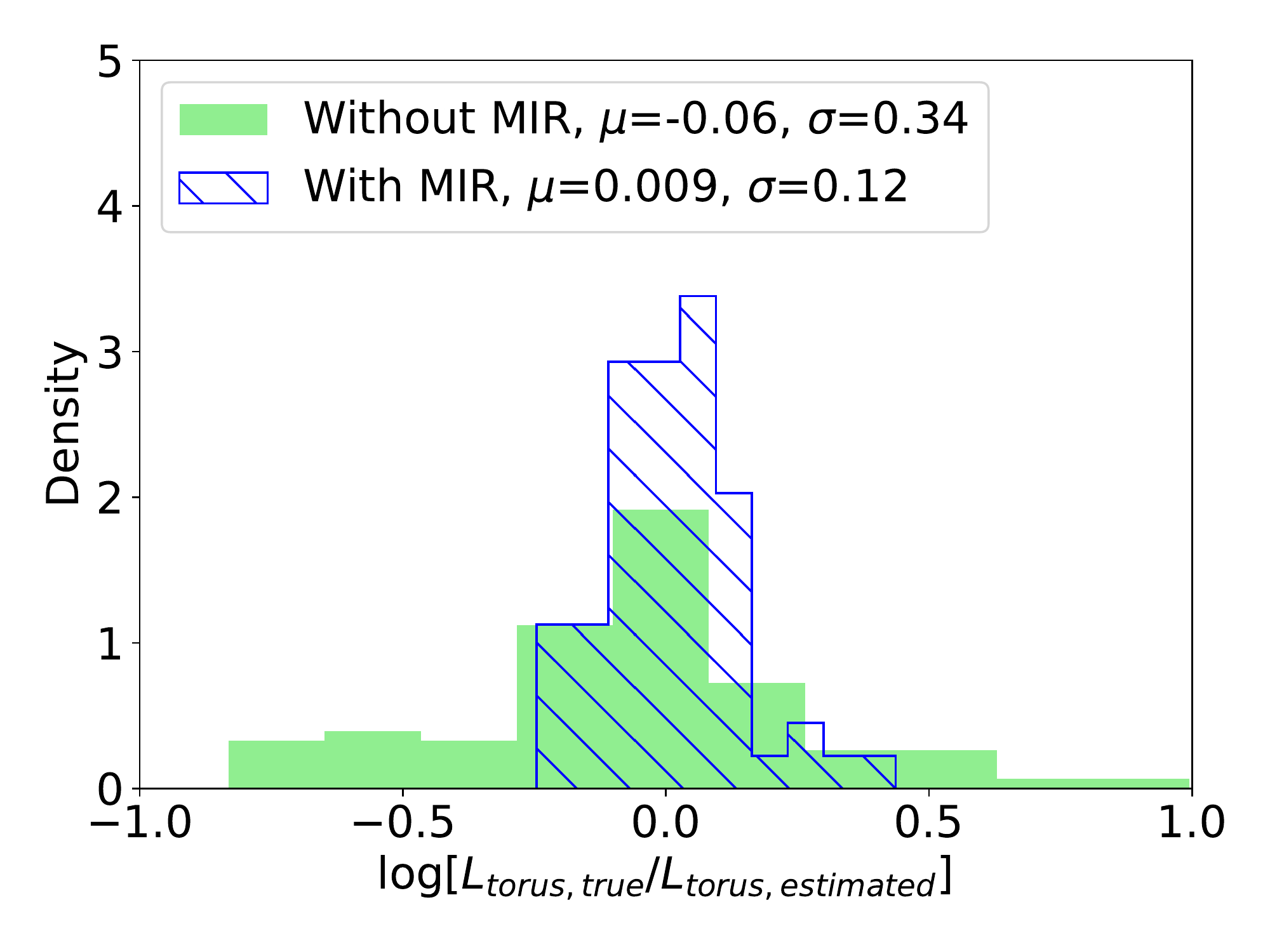} 
    \end{tabular}
\caption{Top panel: Measurements of the $\rm L_{AGN,torus}(erg~s^{-1})$ parameter with and without W4 or MIPS data. The two measurements of $\rm L_{AGN,torus}(erg~s^{-1})$ are in good agreement ($\rm \rho_{torus}=0.800$). The dashed line presents the 1:1 relation. Middle panel: Estimated vs. true values provided by the best-fit model, for $\rm L_{AGN,torus}(erg~s^{-1})$,  for the three fields used in our analysis. The filled (open) symbols represent the X-XIGALE runs with (without) MIR data. The errors on the estimated values increase when the analysis lacks MIR data. The mean value of the error on log[$\rm L_{AGN,torus}(erg~s^{-1})$] is 0.27 dex without MIR, compared to $\sim$0.15 dex when MIR data are used. Bottom panel: Distributions of the difference between true and estimated values, with and without MIR data. The dispersion increases when W4 or MIPS1 are absent, but the estimated $\rm L_{AGN,torus}(erg~s^{-1})$ values are still consistent with the true values.}\label{torusmock}
\end{figure}

In Fig.~\ref{torusmock} (upper panel), we compare the log[$\rm L_{AGN,torus}(erg~s^{-1})$] measurements with and without MIR bands (W4 or MIPS). The two measurements are consistent ($\rm \rho_{torus}=0.800$). In the middle panel of this figure, we present the estimated log[$\rm L_{AGN,torus}(erg~s^{-1})$] versus the true values for the samples with and without MIR photometry, using the mock catalogue. In both cases, we found strong correlation ($\rm \rho_{torus}>0.8$) between the estimated and true values, as indicated in the legend of the plots. We noticed that when we do not include the MIR photometry in the SED fitting, the uncertainties of log[$\rm L_{AGN,torus}(erg~s^{-1})$] are higher compared to the sample with MIR detections. The mean value of the error on log[$\rm L_{AGN,torus}(erg~s^{-1})$] is 0.27 dex in the former case, compared to $\sim$0.15 dex for the latter. The scatter of the log[$\rm L_{AGN,torus}(erg~s^{-1})$] also increases when we ignore the longest MIR photometric bands (bottom panel), but X-CIGALE can still reliably recover the parameter ($\rm \rho_{torus}=0.832$).

\begin{figure}
\center
   \begin{tabular}{c c}
    \includegraphics[width=0.47\textwidth]{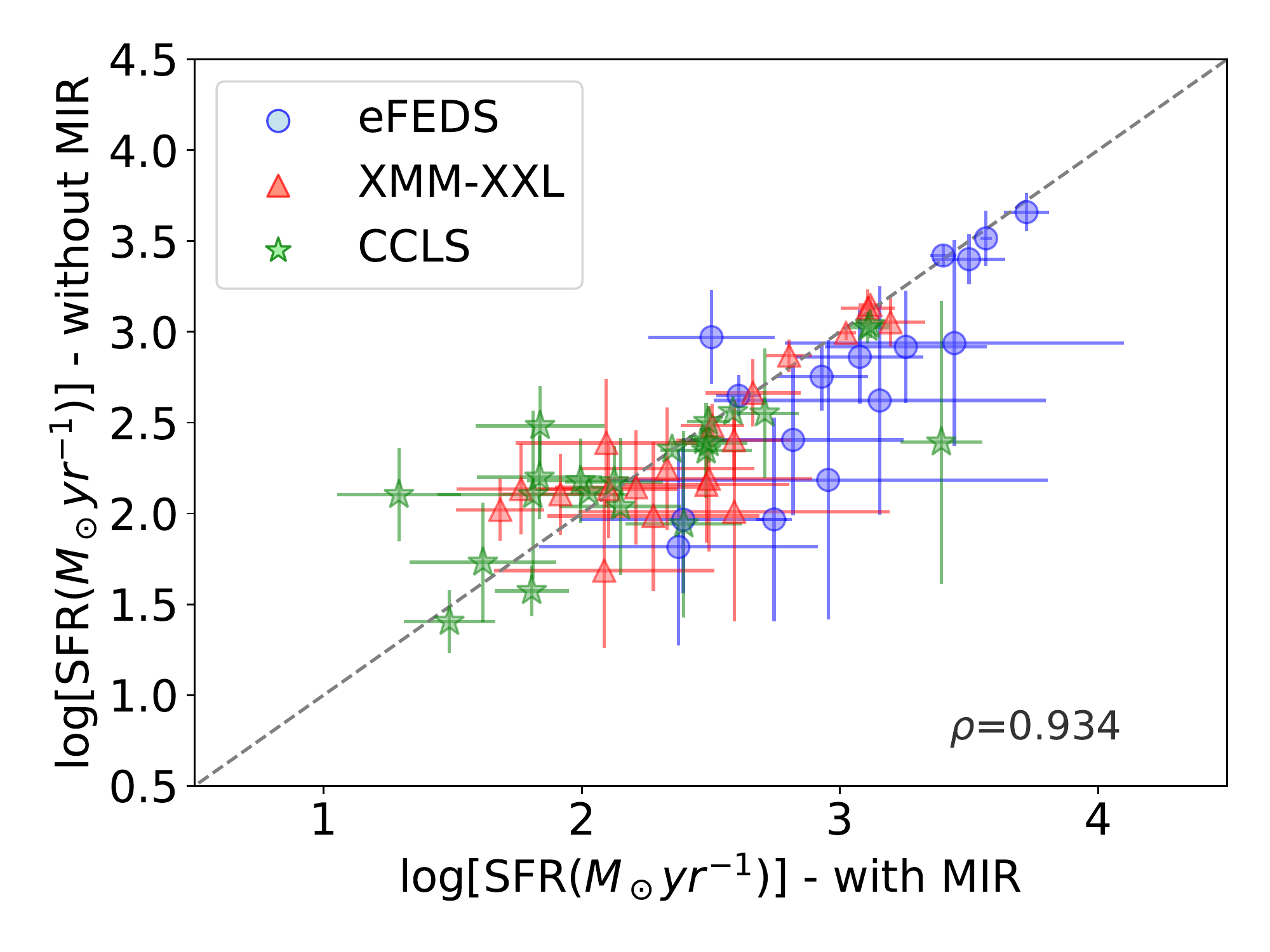} \\
    \includegraphics[width=0.46\textwidth]{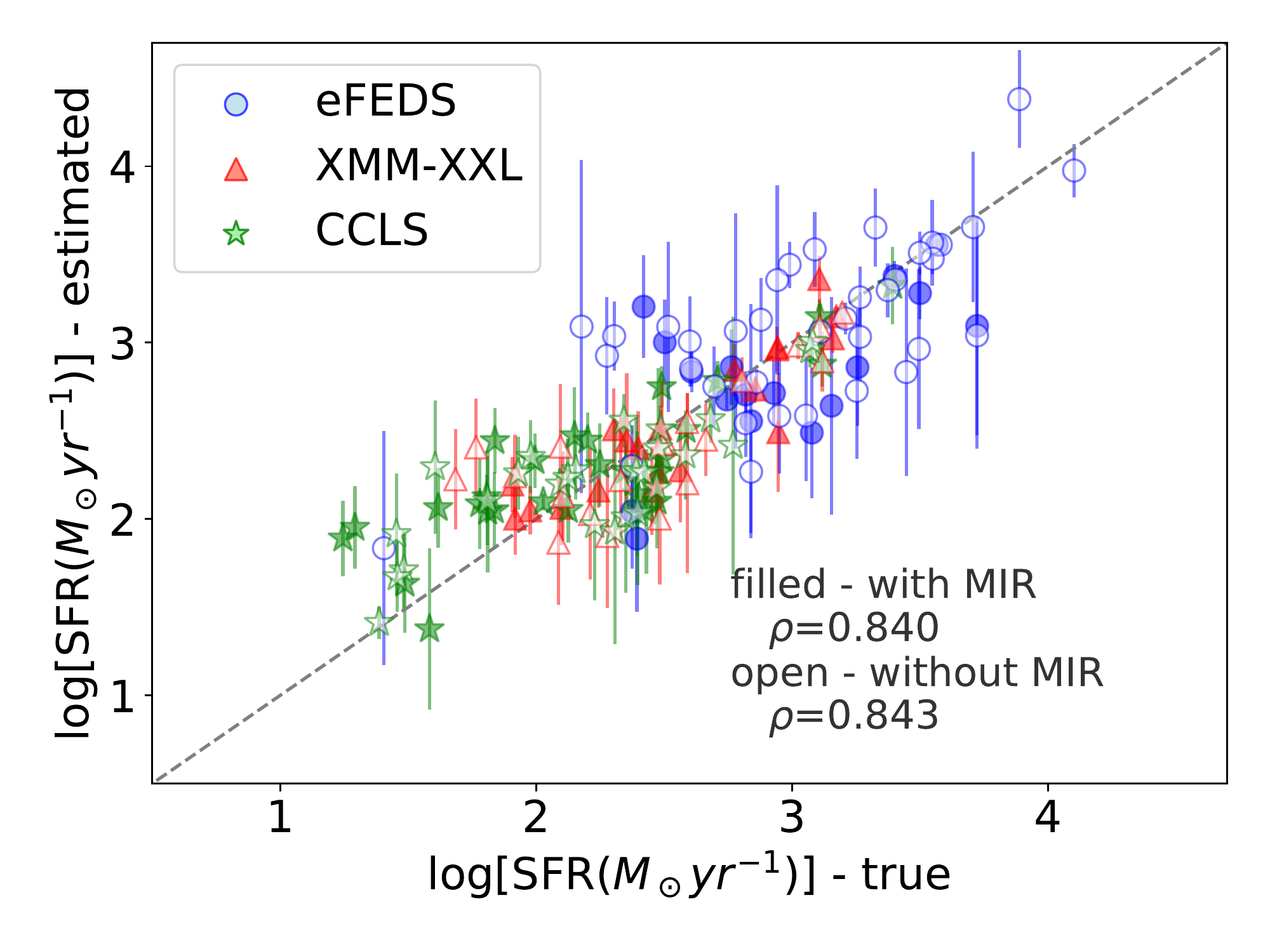} \\
    \includegraphics[width=0.48\textwidth]{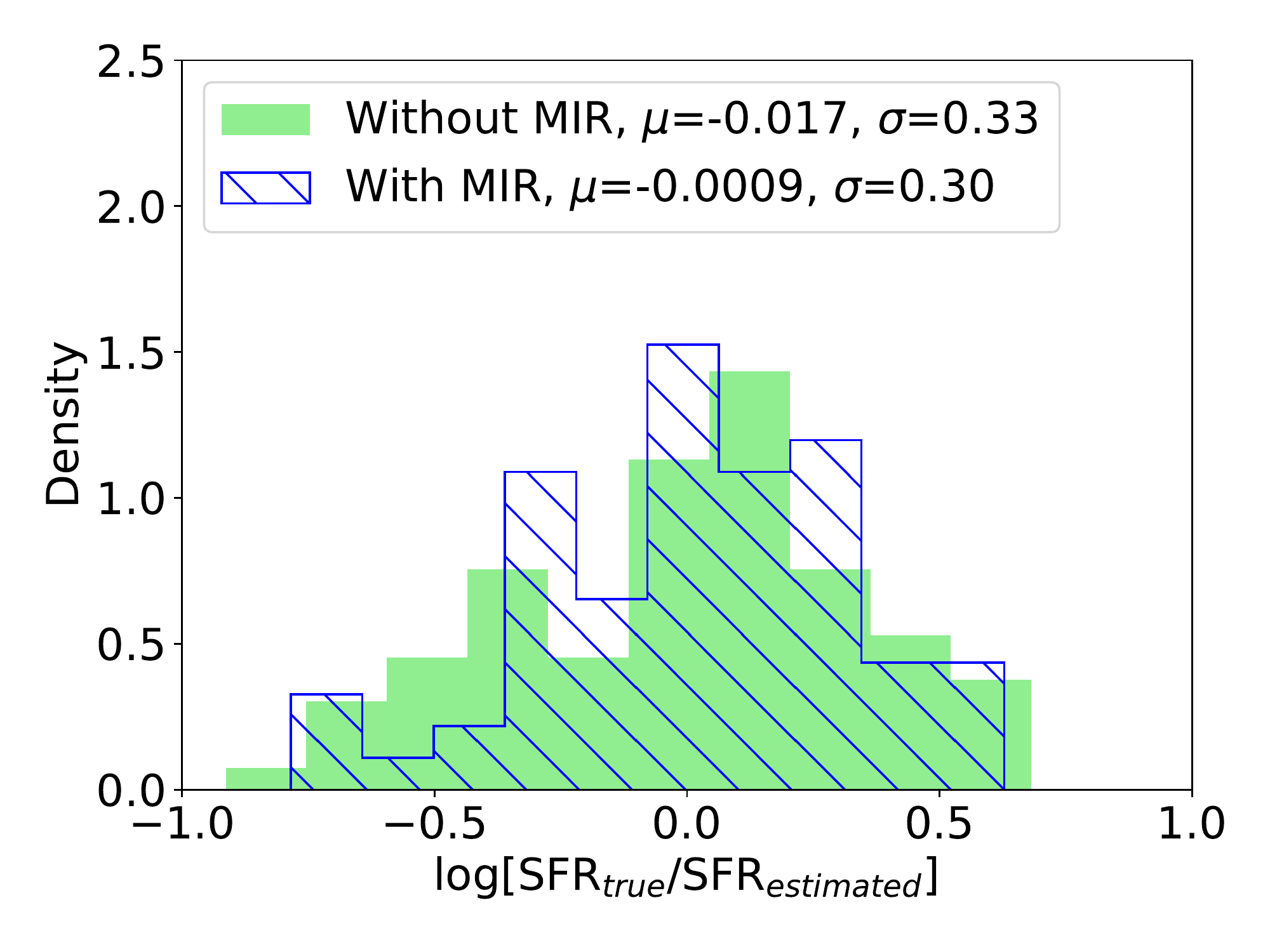}
    \end{tabular}
\caption{Top panel: SFR calculation with and without W4 or MIPS1 photometric data. The dashed line represent the 1-to-1 relation. The two measurements are consistent as indicated by the Pearson coefficient. Middle panel: The estimated SFR based on the mock catalogues compared to the true values provided by the best-fit model. The filled (open) symbols represent the X-XIGALE runs with (without) MIR data. X-CIGALE can successfully recover the SFR of the host galaxy, even in the absense of W4 or MIPS photometry, as indicated by the Pearson correlation factor. Bottom panel: The distributions of the difference between estimated and true values. The two distributions are very similar based on the mean and dispersion values.}\label{mock_results_MIR}
\end{figure}

Finally, we examine whether the lack of the W4 or MIPS1 photometric bands affect the reliability of X-CIGALE to measure the SFR. In Fig.~\ref{mock_results_MIR} (top panel), we present the SFR calculations with and without the W4 or MIPS1 photometric bands, from the SED fitting process. The two SFR measurements are consistent ($\rm \rho_{SFR}=0.934$). In the middle panel, we show the estimated parameters of the host galaxy as functions of the true ones, using the mock analysis. The corresponding linear Pearson correlation coefficients are similar for the two runs indicating not strong deviations in the estimated parameters when these MIR data are missing. In particular, we found $\rm \rho_{SFR}=0.840$ when we include the MIR data and $\rm \rho_{SFR}=0.843$ when we exclude them. Regarding the distribution of the difference between the true and estimated values, in both runs the mean values are very close (bottom panel).

\section{How quality criteria affect the results}\label{appendixA}

\begin{figure}
   \begin{tabular}{c}
       \includegraphics[width=0.47\textwidth]{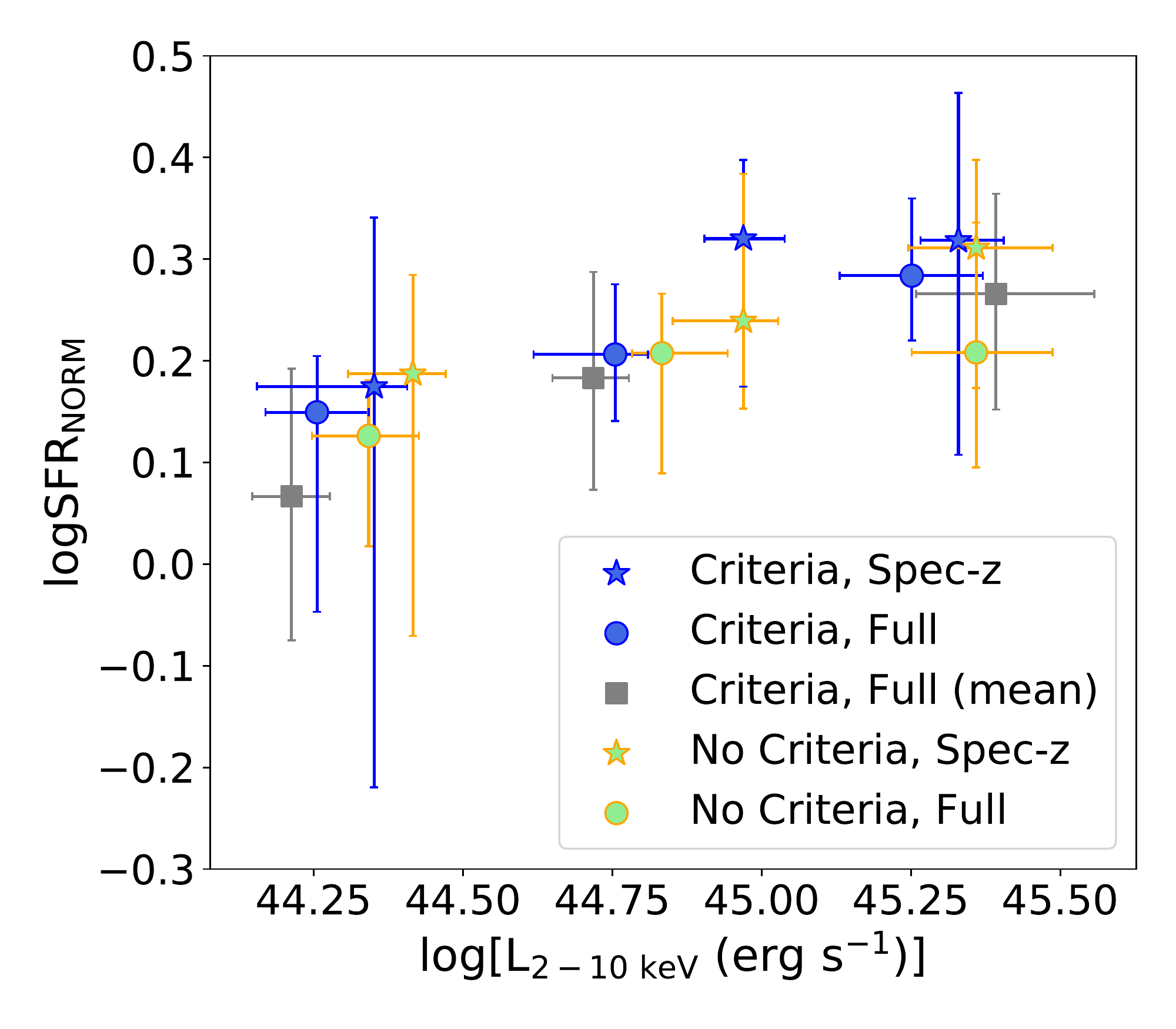} 
    \end{tabular}
\caption{Normalised star formation rate, $\rm SFR_{NORM}$, as a function of rest-frame absorption-corrected X-ray luminosity. Blue (green) colour indicates the samples after (before) applying the quality criteria when including the full sample (circles) and when considering only sources with spectroscopic redshifts (asterisks). In each case, we divided the sample into three bins of equal size and show the median values of $\rm logSFR_{NORM}$ and logL$_X$ with uncertainties calculated using the bootstrap resampling method. The grey squares represent the mean values of the full sample after applying the quality criteria.}\label{sfrnormTEST}
\end{figure}

Out of the 149 initial high-z sources, we used 89 in our analysis with well-fitted SEDs and reliable SFR and stellar mass measurements. 49 out of 89 (55\%) of these sources have available spectroscopic redshifts. In this section, we examined whether 1) including in our analysis only sources with spectroscopic redshifts or 2) using the initial high-z catalogue before applying the selection criteria defined in Sect.\ref{criteria} (e.g., reduced $\chi ^2$, M$_*$ and SFR quality criteria) could affect the results. Following similar analysis as in Sect.~\ref{sec_sfrnorm_lx}, we calculated the median $\rm L_X$ and $\rm log\,SFR_{NORM}$ values of the aforementioned cases and in Fig.\ref{sfrnormTEST} we plot the results. In general, using the sample with only spec-z sources with or without applying the quality selection criteria, we retrieve slightly higher values of $\rm logSFR_{NORM}$. However, our results are in good agreement within the uncertainties.   

Finally, we present the results of the sources that met our selection criteria but instead of the median values we calculated the mean values in each bin (grey squares) with their corresponding uncertainties using the bootstrap resampling method. The resulted binned data of the full sample (both photo-z and spec-z) follow the same trend as in the previous cases. Using mean values for our calculations, instead of median values does not practically affect the $\rm L_X$ values of each bin. However, the bins are shifted towards lower $\rm logSFR_{NORM}$ values. Nonetheless, the data points still lie above the main sequence ($\rm logSFR_{norm}>0$).

\end{appendix}

\end{document}